\documentclass{article}

\usepackage[english]{babel}
\usepackage[letterpaper,top=2cm,bottom=2cm,left=3cm,right=3cm,marginparwidth=1.75cm]{geometry}
\usepackage{mathptmx}
\usepackage{sectsty}
\usepackage{amsmath}
\usepackage{amssymb}
\usepackage[numbers]{natbib}
\usepackage{bm}
\usepackage{graphicx} 
\usepackage{hyperref}

\usepackage{subcaption}

\usepackage[format=plain,justification=justified,singlelinecheck=false,font={stretch=1.125,small,sf},labelfont=bf,labelsep=space]{caption}

\usepackage[T1]{fontenc}
\usepackage[version=4]{mhchem}

\usepackage{newtxmath}

\usepackage{gensymb}

\usepackage{gensymb}
\usepackage{comment}
\usepackage{subcaption}

\title{Experimental and numerical study of the dynamics of sedimenting pairs of semi-flexible fibers close to attractive `aligned' relative configuration
} 

\author{
    Harish~N.~Mirajkar\textit{$^{a}$}$^{,\dagger,\parallel}$, 
    Chandra~Shekhar\textit{$^{a}$}$^{,\ddagger}$, 
    Yevgen~Melikhov\textit{$^{a}$}$^{,\parallel}$, \\
    Piotr~Zdybel\textit{$^{a}$}$^{,\parallel}$, 
    Maria~L.~Ekiel-Je\.zewska\textit{$^{a}$}$^{,\ast}$\\
    {\it \small $^a$Institute of Fundamental Technological Research, Polish Academy of Sciences,} \vspace{-0.1cm} \\{\it \small ul. Pawi\'nskiego 5B, 02-106 Warsaw, Poland}
}

\begin{document}

\renewcommand{\thefootnote}{\fnsymbol{footnote}}
\renewcommand\footnoterule{\vspace*{1pt}
 \hrule width 3.5in height 0.4pt \vspace*{5pt}} 
\newcommand{\bee}{\begin{eqnarray}}
\newcommand{\eee}{\end{eqnarray}}
\newcommand{\ba}{\begin{array}}
\newcommand{\ea}{\end{array}}
\newcommand{\bt}{\begin{tabular}}
\newcommand{\et}{\end{tabular}}

\date{}
\maketitle
\vspace{-0.5cm}
\begin{abstract}
Dynamics of two short semi-flexible elongated objects (called fibers) settling under gravity in a viscous fluid are investigated experimentally and numerically at Reynolds numbers much smaller than unity. 
We focus on fibers initially relatively close to each other, and we check if later they approach an `aligned' horizontal configuration, previously identified numerically (Bukowicki and Ekiel-Je\.zewska, \textit{Soft Matter} \textbf{46} (2019) 9379) as attractive for symmetric initial conditions of moderately elastic filaments. 
In our experiments, two semi-flexible ball chains sediment in a highly viscous silicone oil. 
They are initially straight and close to a parallel horizontal relative configuration, with random deviations. 
Their motion and shape deformation are recorded on a relatively short timescale using two synchronized cameras. 
We find that for most of the trials, relatively short fibers stay together, with damped oscillations around the symmetric `aligned' configuration in which the end-to-end vectors are parallel to each other in a horizontal plane. 
For a few initial conditions, the ball chains tend to move away horizontally or vertically. To study the long-time behavior, we perform numerical simulations, modeling moderately elastic filaments as chains of identical beads, with the centers of consecutive beads connected by springs and with the fibers' elastic resistance to bending. 
Different initial positions and orientations are considered. 
Their dynamics are determined by the multipole expansion of the Stokes equations, implemented in the precise Hydromultipole numerical code. 
For short times, we observe similarity of the dynamics of semi-flexible ball chains and moderately elastic filaments. 
We provide examples of long-time numerical simulations illustrating that elastic filaments close to each other can move away horizontally or vertically, but after a long time, come back again and perform damped oscillations while approaching the aligned configuration with almost touching filament ends. 
We confirm the attractive nature of the aligned configuration of very close semi-flexible sedimenting fibers, even if they are far away from each other.
\end{abstract}


\footnotetext{$^{\parallel}$~Harish N. Mirajkar, Yevgen Melikhov, and Piotr Zdybel contributed equally to this work}

\footnotetext{$^\dagger$~Currently at School of Engineering and Applied Science, Ahmedabad University, Gujarat, India}

\footnotetext{$^\ddagger$~Currently at Department of Chemical Engineering, Thapar Institute of Engineering and Technology, Patiala, Punjab, India-147004}

\footnotetext{$^{\ast}$~Corresponding author: Maria~L.~Ekiel-Je\.zewska 
E-mail: mekiel@ippt.pan.pl}


\section{Introduction}
\label{sec:intro}

Motion and deformation of semi-flexible fibers of micro-meter sizes in viscous environments are significant for biological, medical, and industrial applications.  Cilia and flagella in microorganisms, actin, and artificially produced microfibers are just a few examples \cite{du2019}.

Dynamics of a single elastic filament settling under gravity in a viscous fluid at low Reynolds numbers have been extensively studied theoretically \cite{li_sedimentation_2013}, numerically \cite{schlagberger2005orientation, manghi2006hydrodynamic, lagomarsino,  saggiorato2015conformations, melikhov2024}, and experimentally \cite{marchetti_deformation_2018}. 
It has been shown that stiffer filaments form a stable U-shaped, or unstable W-shaped configuration \cite{lagomarsino,  saggiorato2015conformations, marchetti_deformation_2018, melikhov2024, llopis_sedimentation_2007}. 
More elastic filaments attain one of several complex attracting dynamical modes \cite{saggiorato2015conformations}, depending on their aspect ratio and the ratio of gravitational to elastic bending forces \cite{melikhov2024}. 
Highly elastic filaments can rotate and move sideways, keeping a fixed shape or performing time-dependent shape deformation, in a periodic or quasi-periodic pattern \cite{saggiorato2015conformations, melikhov2024}. 
The influence of relatively small Brownian motion has also been studied \cite{cunha2022settling}.

The natural extension is to study dynamics of pairs of elastic filaments. The basic questions are whether, and under what conditions, the fibers approach or move away from each other, whether they synchronize their orientation while translating, rotating, and performing periodic or quasi-periodic oscillations of positions, orientations, or shapes, and whether there exist stationary attracting relative configurations.

Numerical investigation of a sedimenting pair of elastic filaments has so far been focused on several symmetric initial configurations. 
It was shown numerically that initially horizontal moderately elastic filaments, located one above the other in the same vertical plane, later decrease their distance \cite{lagomarsino,saggiorato2015conformations}. 
If in the above initial configuration one of the elastic filaments is rotated about the vertical axis $z$, then later both filaments rotate around $z$ with different angular velocities; they tend to the same vertical plane, and approach each other \cite{saggiorato2015conformations}. 
Moderately elastic filaments, initially placed in a horizontal co-linear configuration, perform a tumbling motion \cite{llopis_sedimentation_2007}.

It has been numerically shown~\cite{bukowicki_sedimenting_2019} that two elastic filaments sedimenting under gravity in a very viscous fluid, initially at the mirror-symmetric orientations at a certain angle in a horizontal plane, after a relatively short time of periodic oscillations of positions and orientations, approach an `aligned' configuration in which their end-to-end vectors are horizontal and parallel to each other. They keep such orientations for a long time. 
Moreover, stiffer or shorter fibers at the aligned configuration come close to each other, almost touching at their ends ~\cite{bukowicki_sedimenting_2019}. This hydrodynamic attraction, and the resulting long-time proximity of different filaments, might be used to trigger some chemical reactions, or biological processes.

Therefore, it is important to explore if the aligned configuration is stable against random perturbations, 
and what are the characteristic times to approach the `aligned' configuration, and to reach a close proximity of semi-flexible fibers. This work is aimed at getting some understanding of these issues, using both experimental and numerical approaches, but with the main focus on the experiments.

In the experiments, we model the semi-flexible fibers by relatively short ball chains. This choice is motivated by a similar approach in the literature. Namely, it has been shown that experiments with semi-flexible knotted loops made of ball chains are consistent with the numerical simulations of the knotted loops made of elastic filaments \cite{gruziel2018}. The experiments with sedimenting open ball chains reproduced well the numerical simulations of elastic filaments; in particular, the unstable W-shaped configuration of a ball chain was detected experimentally \cite{shashank2023dynamics}.

In this study, we experimentally investigate the dynamics and interactions of a pair of semi-flexible ball chains settling under gravity in a highly viscous fluid. 
We initially put the two ball chains approximately parallel and close to each other in the same horizontal plane. We observe how their positions and orientations change with time, and whether the system tends towards the aligned configuration. 
In the complementary numerical simulations of elastic filaments, we aim to reproduce the measured short-time dynamics, and also to get insight into how the system evolves on a much longer timescale.

The paper is structured as follows. 
Section~\ref{sec:exp_meth} describes the experimental setup, the details of the flexible ball chains, and the image processing technique. 
Section~\ref{sec:theor_num_approach} explains the theoretical and numerical approach. 
In section~\ref{sec:res_discuss}, we present the experimental results, focusing on the evolution of ball chains of varying length (3 to 6 beads). 
In section~\ref{sec:num}, we discuss the numerical simulation results obtained using {\sc Hydromultipole} numerical codes. 
Finally, section~\ref{sec:concl} provides the conclusions.

\section{Experimental methodology}
\label{sec:exp_meth}

\subsection{Experimental setup}
\label{sec:exp_meth:exp_setup}

A glass tank with internal dimensions of 200~mm in width, 200~mm in depth, and 500~mm in height is filled up to a depth of 495~mm with highly viscous silicone oil (manufactured by Silikony Polskie) with a kinematic viscosity of $5 \times 10^{-3}$~$\text{m}^{2}/\text{s}$ and a density of 970~$\text{kg}/\text{m}^{3}$.
In this tank, we release a pair of flexible metallic ball chains, whose motion in the viscous fluid is of primary interest in this study. The initial positions of the ball chains are on the surface of the fluid and in the central part to minimize hydrodynamic interactions with the side walls. 
To initiate the motion, a tip of each ball chain is positioned within one of two short parallel grooves separated by approximately 7~mm at the end of the tweezers. 
Next, the tweezers are used to position the ball chains in parallel at the free surface of the oil. 
Then, the tweezers are opened.

\begin{figure}[h!]
\centering
    \includegraphics[width=0.5\linewidth]{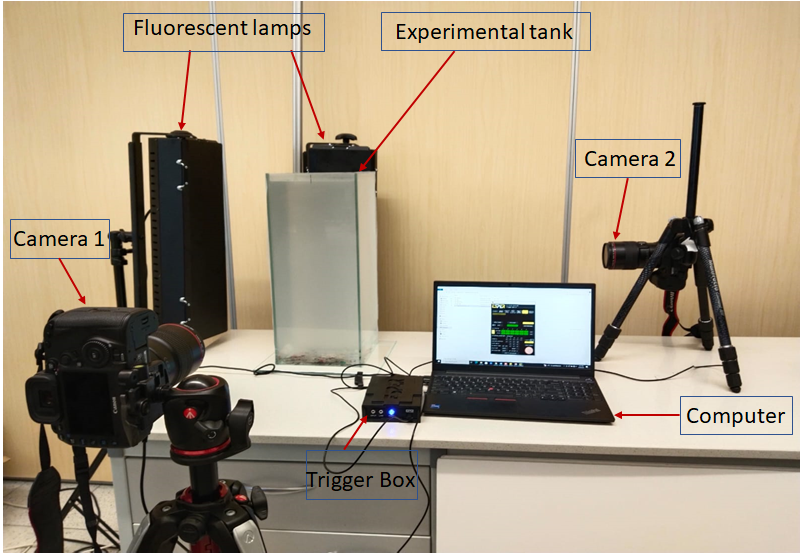} \vspace{-0.2cm}
 \caption{The experiment setup.
    }
    \label{fig:exp_setup}
\end{figure}

To capture the behavior of the ball chains with high precision, two Canon 5D Mark IV DSLR cameras, each with a 100~mm prime lens, are used to provide synchronized front and side views of the tank (as shown in Fig.~\ref{fig:exp_setup} and Refs.~\cite{shashank2023dynamics, shekhar2026}).  
Camera~1 directly faces the front wall of the tank, capturing a front view, while Camera~2 faces the side wall of the tank, capturing a side view (see Fig.~\ref{fig:exp_setup}).~These cameras, which offer a resolution of 30 megapixels, are positioned perpendicularly to each other to collect visual information on lateral and vertical 3D movements, as well as the instantaneous shapes of the ball chains and their 3D orientation. 
The tank is illuminated by two fluorescent lamps placed behind the tank walls opposite to the cameras (see Fig.~\ref{fig:exp_setup}). 
Tracing paper is attached to these walls to diffuse the light. 
This backlight design provides bright, uniform lighting that clearly outlines the dark, opaque ball chains against the bright background, making their motion easily distinguishable in the captured images, as shown by \cite{shashank2023dynamics}. 

Both cameras are synchronized using a Trigger Box (manufactured by Esper), which allows precise timing with a delay of less than 20~ms between the two camera triggers, ensuring simultaneous capture by the cameras from both perspectives. 
The Trigger Box is controlled by a laptop with software, enabling continuous capture at the cameras with a frame rate of 1 image per second. 
Each image is captured with an exposure time of 1/125~s. 
The movement of the ball chains during this interval is smaller than half of a pixel. 
To maintain depth of field and image sharpness throughout the descent of the chains, the $f$-number is set to $f$/32 (the smallest aperture available), and the ISO is kept at 160 to balance the brightness of the image with low noise. 
The cameras are positioned in portrait orientation at a distance of approximately 820~mm from the front and side surfaces of the tank.
This setup provides a field of view centered within the tank, covering approximately 350~mm of the fluid in the vertical direction, and including the whole tank width, i.e., 200~mm in the horizontal direction.  
Intentionally, about 75~mm from both the top and bottom of the tank are excluded from the captured images to avoid the influence of the free surface at the top and the wall at the bottom on the recorded dynamics of the sedimenting ball chains.

\subsection{Ball chains}
\label{sec:exp_meth:ball_chain}

\begin{figure}[h!]
    \centering
    \begin{subfigure}{0.19\textwidth}
        \centering
        \includegraphics[width=\textwidth]{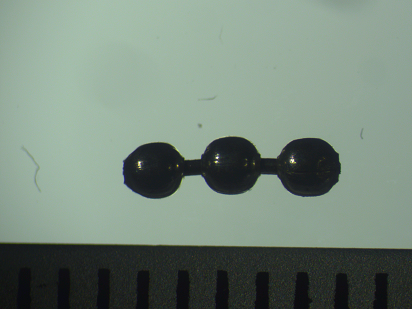}
        \caption{$N$ = 3}
        \label{fig:ballchains:a}
    \end{subfigure}
    \hspace{0.5cm}
    \begin{subfigure}{0.19\textwidth}
        \centering
        \includegraphics[width=\textwidth]{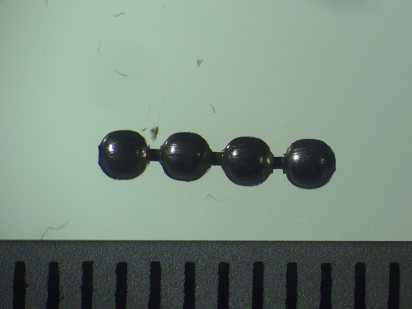}
        \caption{$N$ = 4}
        \label{fig:ballchains:b}
    \end{subfigure}\\
    \begin{subfigure}{0.19\textwidth}
        \centering
        \includegraphics[width=\textwidth]{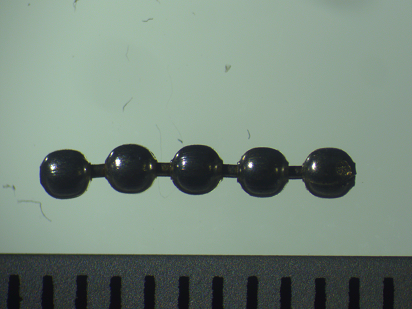}
        \caption{$N$ = 5}
        \label{fig:ballchains:c}
    \end{subfigure}
    \hspace{0.5cm}
    \begin{subfigure}{0.19\textwidth}
        \centering
        \includegraphics[width=\textwidth]{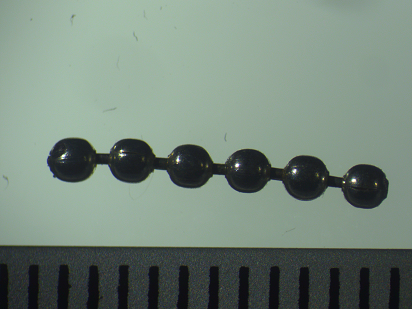}
        \caption{$N$ = 6}
        \label{fig:ballchains:d}
    \end{subfigure}
    \vspace{-0.3cm}
    \caption{Photographs of the ball chains used in the experiments. 
        Consecutive ticks on the scale are separated by 1 mm.
    } \vspace{-0.2cm}
    \label{fig:ballchains}
\end{figure}

To model the dynamics of elastic fibers, we use semi-flexible ball chains, following the idea presented by \citet{gruziel2018} and applied by \citet{shashank2023dynamics}.
The long ball chains used in the experiments are obtained from Manzuko, an online store (product no LL01114SS).  They consist of hollow metallic beads with a diameter $d=1.5$~mm coated with a silver color. These beads are linked in series by rigid metallic dumbbells that pass through central holes in each bead. There is a slack between the middle part of the dumbbell and the edges of the hole in a bead. This configuration allows for small changes in the distance between consecutive beads and significant bending of the ball chain, with the bending angle not exceeding $52^{\circ}$. 
We cut the long ball chains to the required length. 
Ball chains of different lengths (i.e., different numbers of beads $N$) have been used,  with $N = 3,4,5,\text{and } 6$ (see Fig.~\ref{fig:ballchains}).

\subsection{Image analysis and calibration}
\label{sec:exp_meth:image_anal}

To analyze the motion of the ball chains, we extract relevant data from sequential photographic images captured by both cameras. Each image has a size of 6720 (height) x 4480 (width) pixels. These images are imported into MATLAB, where image-processing techniques are applied to identify the position of each bead from both chains. The processing begins with image thresholding and binarization, which convert the images into a binary format to distinguish the ball chains from the background. Noise removal and gap-filling techniques are subsequently applied to eliminate inconsistencies caused by shadows or reflections, ensuring that each ball chain is distinctly identified in every frame and from both camera views.

To convert image coordinates from pixels to millimeters, we performed a spatial calibration using a standard scale positioned along the central vertical axis of the fluid tank. 
The field of view was adjusted such that Camera 1 and Camera 2 captured an identical vertical span.
The calibration was performed before each of the two experimental sessions. 
As a result, the effective image height, measured in millimeters, varied slightly across trials from different sessions. 
In the following, we will present the results from 7 experimental trials, labeled as $m_1$, ..., $m_7$ (all the trials are shown in the repository \cite{repository}). 
For trials $m_1$, $m_2$, $m_5$ and $m_7$, the 6720-pixel span corresponded to 345~mm, yielding the calibration factor 0.0513~mm/pixel. 
For trials $m_3$, $m_4$ and $m_6$, the same pixel span corresponded to 350~mm, yielding the calibration factor 0.0521~mm/pixel.
In both cases, the calibration was assumed to be isotropic, and the same factors were applied to horizontal displacements for the respective datasets.
These calibration factors were applied to determine the positions of the ball chains close to the central vertical axis of the container.

\subsection{Extracting parameters of the dynamics from the binary images}
\label{sec:exp_meth:pa}

We analyze time-ordered, binarized 2D images of ball chains using a semi-automated workflow that combines circle detection with user-guided chain assignment. Image processing and geometric calculations are performed with the Python library \texttt{OpenCV} for Hough transforms and \texttt{NumPy} for calculations on vectors. The procedure is described in detail in Appendix \ref{appendix_bead_detection}.

\begin{figure}[h!]\vspace{-0.1cm}
    \centering 
    \begin{subfigure}{0.4\textwidth}
        \centering
\includegraphics[width=0.89\textwidth]{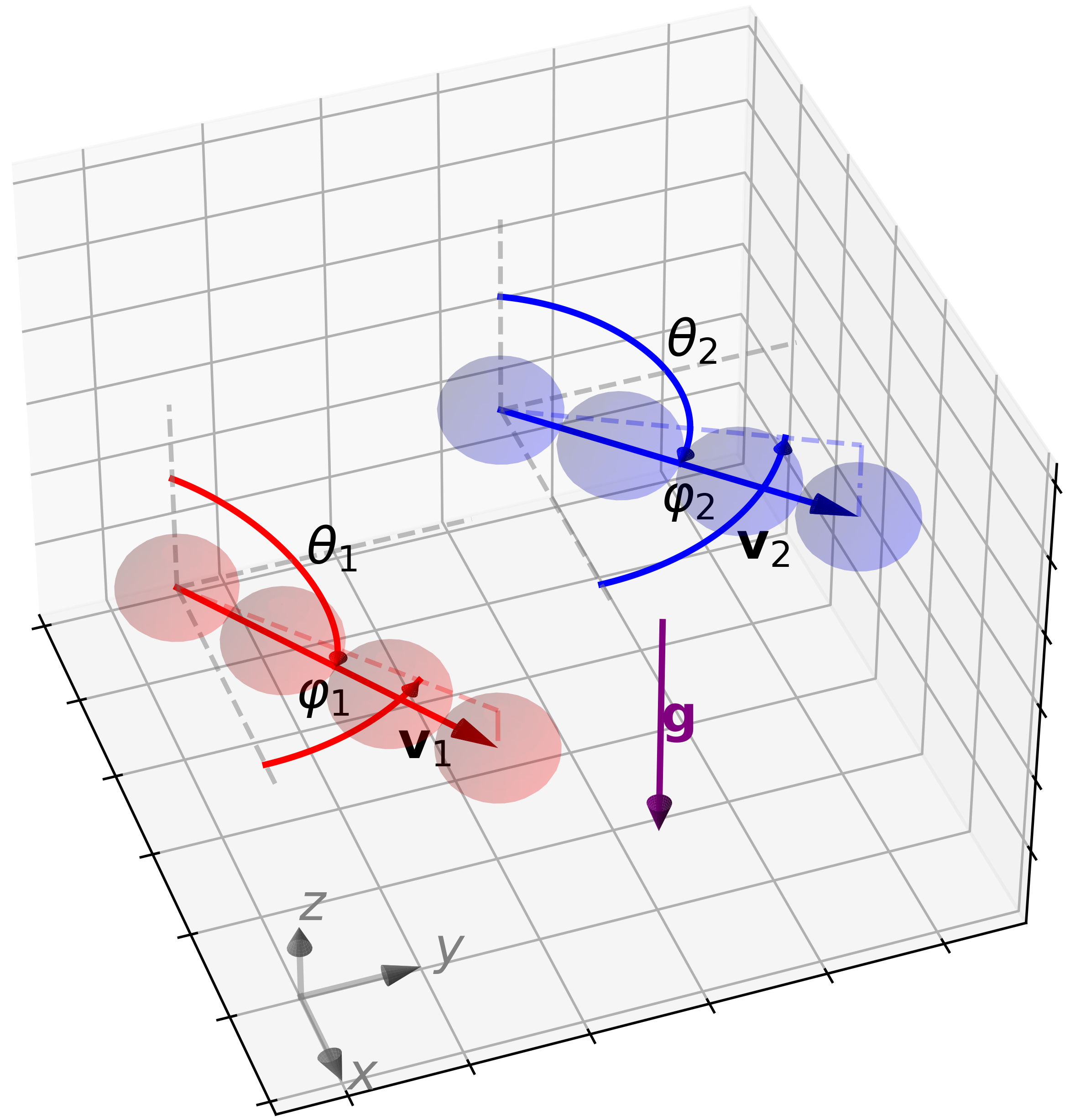}
        \vspace{-0.2cm}
       \caption{}
        \label{fig:notation:a} \vspace{-0.1cm}
    \end{subfigure} 
    \begin{subfigure}{0.4\textwidth}
        \centering
        \includegraphics[width=0.8\textwidth]{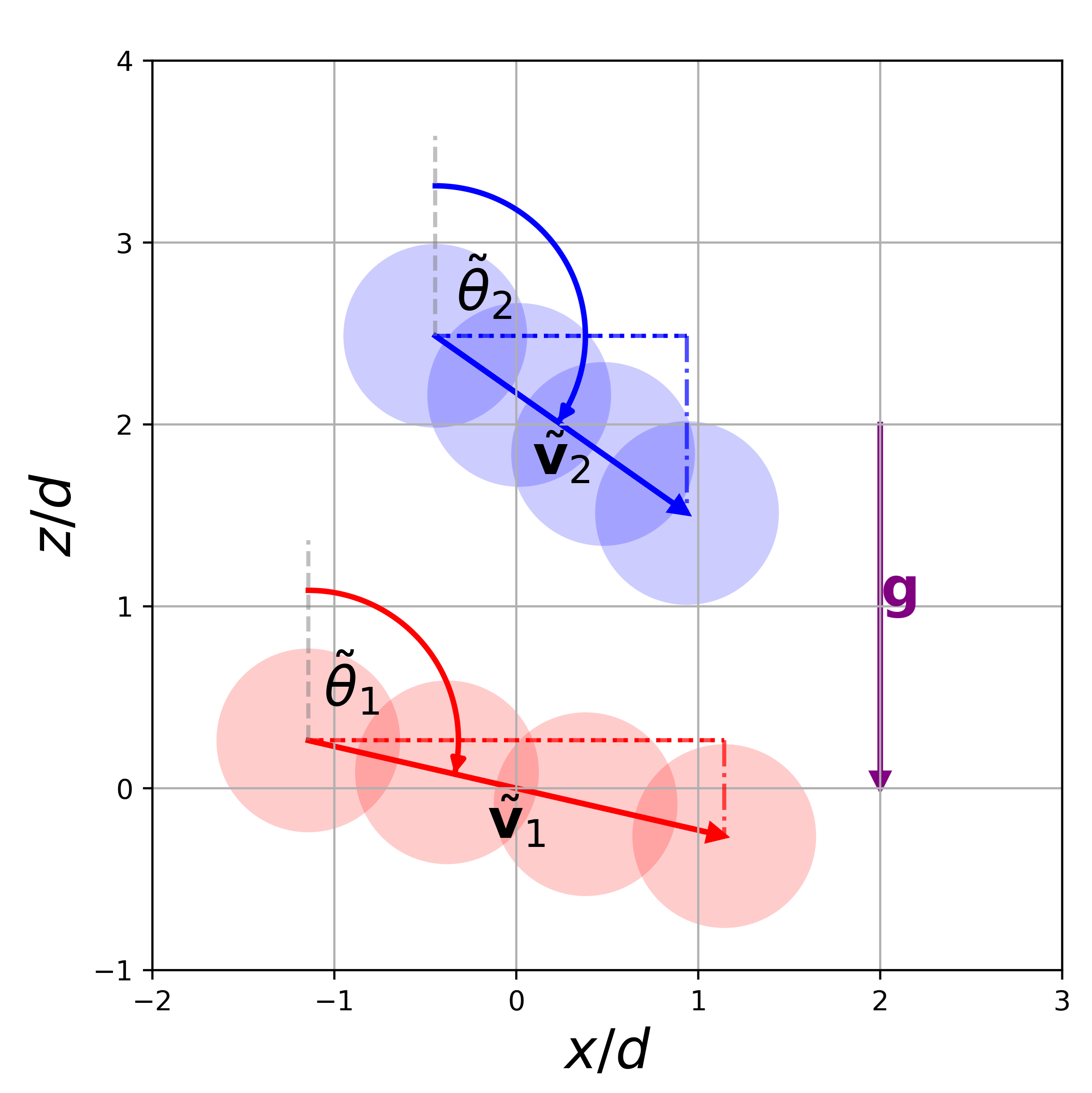}%
        \vspace{-0.25cm}
        \caption{}
        \label{fig:notation:b}
    \end{subfigure} \vspace{-0.2cm}
    \caption{
        The parametrization of the orientation of elastic fibers (or ball chains) made of beads with diameter $d$ :
        (a) by  $\tilde{\theta}_{i}$ in 2D projection onto the $xz$‐plane; 
        (b) by $\theta_{i}$, $\varphi_{i}$ in 3D space. 
        Gravitational acceleration $\mathbf{g}$ is anti‐parallel to the $z$‐axis.}
    \label{fig:notation}
\end{figure}

Let frames be indexed by $f\in\mathcal{F}$, which is a subsampled subset of the full sequence of frames obtained by uniform temporal subsampling of the original Camera~1 image sequence. 
More precisely, if $\Delta f=2$ denotes the subsampling step in frame number, then the analyzed set has the form $\mathcal{F}=\{f_0,f_0+\Delta f,f_0+2\Delta f,\dots\}$ within the selected frame interval. 
For each processed frame $f$ from Camera~1, the method detects the beads in the corresponding binary image using primarily a circular Hough-transform procedure constrained to a narrow radius range around a common expected bead radius. 
Thus, the method effectively fits circles of approximately equal diameter to the bead images, allowing only a small tolerance around the nominal radius. 
The bead centers are then estimated as the centers of the detected circles. 
If the Hough-based detection is insufficient, a conservative contour-based fallback is used, in which the bead center is taken as the centroid of the detected contour. 
Hence, $(x_{i},z_{i})$ denote the coordinates of the detected bead centers in the adopted laboratory-frame convention. 
The procedure then requests to manually select ordered bead subsets that form each chain. 
Then, for each chain, the time-dependent center-of-mass position, denoted by $(x_{c1},z_{c1})$ and $(x_{c2},z_{c2})$, is computed as the arithmetic mean of the coordinates of the selected bead centers. 
In addition, the two-dimensional end-to-end vectors $\tilde{\mathbf v}_1$ and $\tilde{\mathbf v}_2$ are defined as the vectors connecting the centers of the first and last selected beads in each chain. 
Angles $\tilde{\theta}_{i}$ between the gravity direction $z$ and $\tilde{\bf v}_{i}$ are then determined for each chain $i$, consistent with the adopted coordinate convention, as illustrated in Fig.~\ref{fig:notation:a}.

To determine the distance between the ball chains, we also apply the same method to images from Camera 2, and the time-dependent positions $(y_{ci},z_{ci})$ of the center of mass of each ball chain $i$ are determined. Then, we evaluate vertical and horizontal separation between the fibers as 
\begin{equation}
    \!\!\!\Delta Z\!=\!z_{c1}\!-\!z_{c2}, \;\, 
          \mbox{and}
    \;\;\Delta H\!=\!\sqrt{(x_{c1}-x_{c2})^2+(y_{c1}-y_{c2})^2}. \;\;\;
\label{definition_of_Deltas}
\end{equation}

\section{Theoretical and numerical approach}
\label{sec:theor_num_approach}

We model each of two elastic fibers as a chain of $N$ identical spherical beads of diameter $d$. The centers of consecutive beads in a chain are connected by springs of the equilibrium length ${l}_0=1.01d$.
The stretching potential energy of the whole fiber has the form,
\begin{equation}
    U_{s} = - 
    \frac{1}{2} k (l_{0}-d)^{2}
    \sum_{n=1}^{N-1} 
    \ln\left[ {1 - \left( \frac{l_{0}-l_{n}}{l_{0}-d} \right)^{2}} \right], 
\end{equation}
where ${{l}}_n=|{\bm{r}}_{n+1}\!-\!{\bm{r}}_{n}|$ is the time-dependent distance between the centers of beads $n\!+\!1$ and $n$, located at $\bm{r}_{n+1}$ and ${\bm{r}_{n}}$, respectively.

In the elastic equilibrium, the fiber is straight. 
Triplets of consecutive beads resist bending, with the bending potential energy of the whole fiber equal to
\begin{equation}
    U_b\!=\!\frac{{A}}{2{l}_0}\sum_{n=2}^{N-1} \beta_n^2,
\end{equation}
where $\beta_n$ is the bending angle,  
$\cos \beta_n\!=\!({\bm{r}}_n-{\bm{r}}_{n-1})\cdot({\bm{r}}_{n+1}-{\bm{r}}_{n})/({l}_n {l}_{n+1})$. 
The spring constant $k$ and bending stiffness $A$ are proportional to the Young modulus $E$ \citep{bukowicki_sedimenting_2019, bukowicki_different_2018},
\begin{equation}
    k = \frac{\pi E d^2}{4 l_{0}}, \hspace{1cm}   A = \frac{E \pi d^4}{64}.
\end{equation}

The total external force $\bm{F}_n$ acting on the $n$-th bead is the sum of the elastic force, 
$\bm{F}_n^e = - \partial \left( U_{s} +  U_{b} \right)/\partial \bm{r}_n$, 
and gravitational force, corrected for buoyancy,  
$\bm{F}^g =-mg\bm{ \hat{e} }_z$,  
where $m$ is the mass of the bead minus the mass of the fluid of the same volume, $g$ is the value of the gravitational acceleration, and  $\bm{ \hat{e} }_z$ is the unit vector along the $z$-axis.

In our system, the Reynolds number is much smaller than unity. 
Therefore, we assume that the fluid flow obeys the Stokes equations, and velocities of the bead centers, $\dot{\bm{r}}_{n}$, depend linearly on the total external forces $\bm{F}_m$ exerted on the beads $m$ from both fibers, 
\begin{equation}
    \dot{\bm{r}}_{n} = \sum_{m=1}^{2N} \bm{\mu}_{nm}(\bm{r}_1,...,\bm{r}_{2N}) \cdot  \bm{F}_m,\label{Sd}
\end{equation}
where the 3x3 mobility matrices $\bm{\mu}_{nm}(\bm{r}_1,...,\bm{r}_{2N})$, depend on the time-dependent positions $\bm{r}_k$ of all the bead centers $k=1,...,2N$ from both fibers, and are evaluated by the multipole expansion of the solutions to the Stokes equations, corrected for lubrication to speed up the expansion convergence \citep{Felderhof1988, Cichocki1994, cichocki_lubrication_1999, ekiel-jezewska2009}.   


The main parameters of the time-dependent relative configuration of fibers (or ball chains) $i\!=\!1,2$ are the 3-dimensional positions of their centers of mass, $(x_{ci},y_{ci},z_{ci})$, and the 3-dimensional end-to-end vectors $\mathbf{v}_i$ connecting the centers of the first and last beads. 
The angle $\theta_i$ is measured from the $z$‐axis to $\mathbf{v}_i$, while the angle $\varphi_i$ is measured from the $x$‐axis to the projection of $\mathbf{v}_i$ onto the $xy$‐plane, as illustrated in Fig.~\ref{fig:notation:b}.

Note that in the simulations, since we have all the 3-dimensional positions of the bead centers, we use angles $\theta_i$, which are different from angles $\tilde{\theta}_i$.
The angle $0 \le \tilde{\theta}_i\le \pi$ is an experimentally measured angle in the plane of view of Camera~1. 
$\tilde{\theta}_i$ is the angle from the $z$-axis to the vector $\tilde{\mathbf{v}}_i$ that is the projection of $\mathbf{v}_i$ on the $xz$-plane, see Fig.~\ref{fig:notation:a}. 
Therefore, 
$\tilde{\theta}_i=\mathrm{atan}(\sin\theta_i \cos\varphi_i/\cos\theta_i)$ 
  for
    $\cos\varphi_i \cos\theta_i>0$, 
$\tilde{\theta}_i=\pi +\mathrm{atan}(\sin\theta_i \cos\varphi_i/\cos\theta_i)$ 
  for 
    $\cos\varphi_i \cos\theta_i<0$,
$\tilde{\theta}_i=\pi/2$ 
  for 
    ${\theta}_i=\pi/2$, and 
$\tilde{\theta}_i$ is not defined 
  for 
    $\cos\varphi_i=0$.

\section{Experimental results}
\label{sec:res_discuss}

We performed 56 experimental trials with pairs of identical ball chains made of $N=3,4,5,6$ beads, released at the surface of the fluid approximately straight and close to a parallel and approximately horizontal relative configuration. 
The distance between the centers of mass of the ball chains was around 7~mm.

The main goal was to check if during the evolution, the ball chains would stay close to each other, and close to the relative configuration, called in brief ``aligned'' (as in Ref.~\cite{bukowicki_sedimenting_2019}), defined by the following conditions:
\begin{equation}
    \mathbf{v}_1 \!\parallel \!\mathbf{v}_2, \;\;\mathbf{v}_i \cdot  \bm{ \hat{e} }_z\!=\!0, \;\;
    \mathbf{v}_i \cdot (\bm{r}_{c1}\!-\!\bm{r}_{c2})\!=\!0, \;\;  (\bm{r}_{c1}\!-\!\bm{r}_{c2}) \cdot \bm{ \hat{e} }_z\!=\!0,\;\;\label{aligned}
\end{equation}
where $\bm{r}_{ci}=(x_{ci},y_{ci},z_{ci})$ is the center-of-mass position of ball chain $i$. 

In the experiments, the range of settling velocities was 2.1-2.6 mm/s, with lower values corresponding to $N\!=\!3$, and larger ones to $N\!=\!6$. 
The corresponding Reynolds number based on the ball chain width (the bead diameter) was Re = $(6.3-7.8)\cdot 10^{-4}$. 

\begin{figure*}
   \centering
   \begin{subfigure}{0.84\textwidth}
        \centering
    \textbf{(a)} \includegraphics[clip, width=0.99\textwidth]{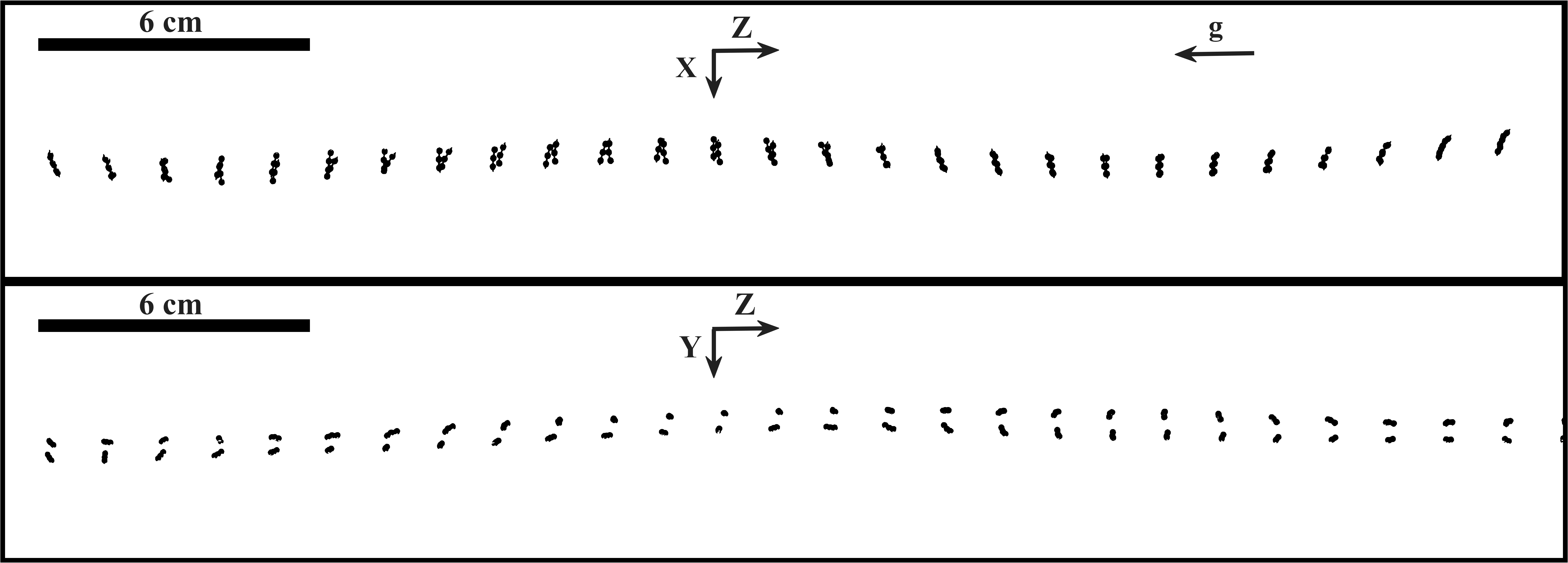}
        \phantomsubcaption
        \label{fig:allmodes:a}
        \vspace{0.25cm} 
    \end{subfigure}
    \begin{subfigure}{0.84\textwidth}
        \centering
    \textbf{(b)} \includegraphics[clip, width=0.99\textwidth]{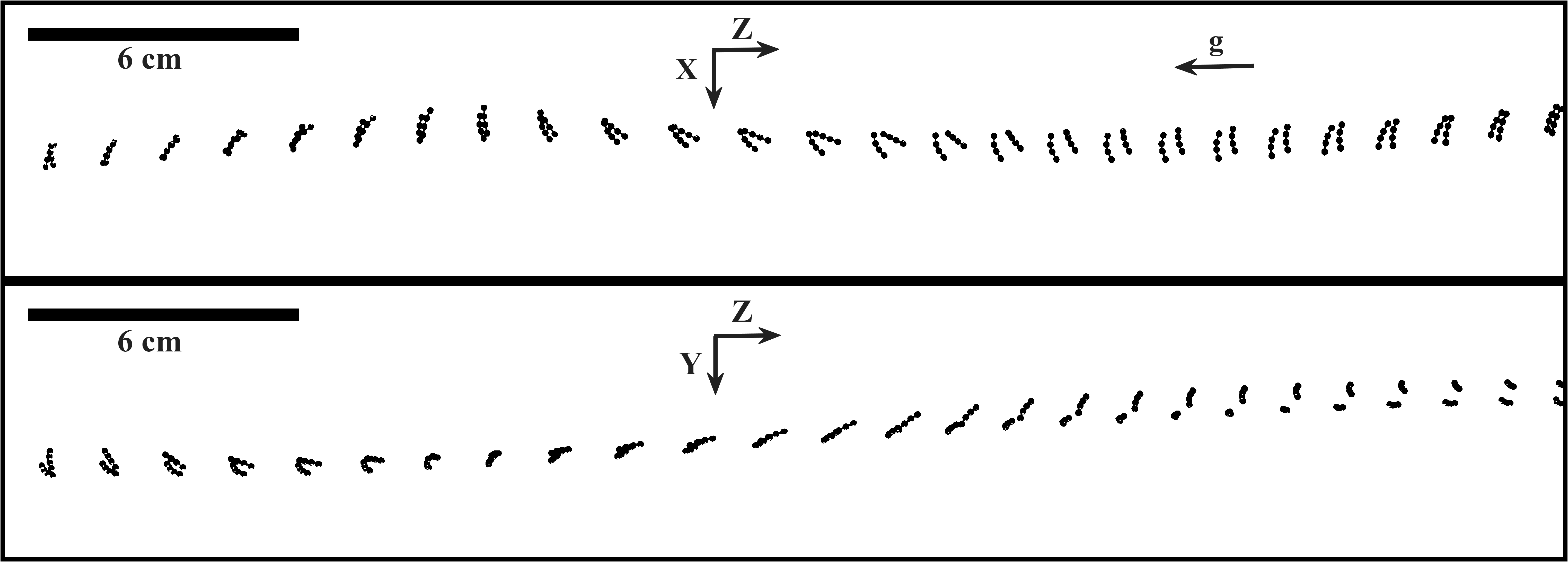}
        \phantomsubcaption
        \label{fig:allmodes:b}
        \vspace{0.25cm} 
    \end{subfigure}
    \begin{subfigure}{0.84\textwidth}
        \centering
    \textbf{(c)} \includegraphics[clip, width=0.99\textwidth]{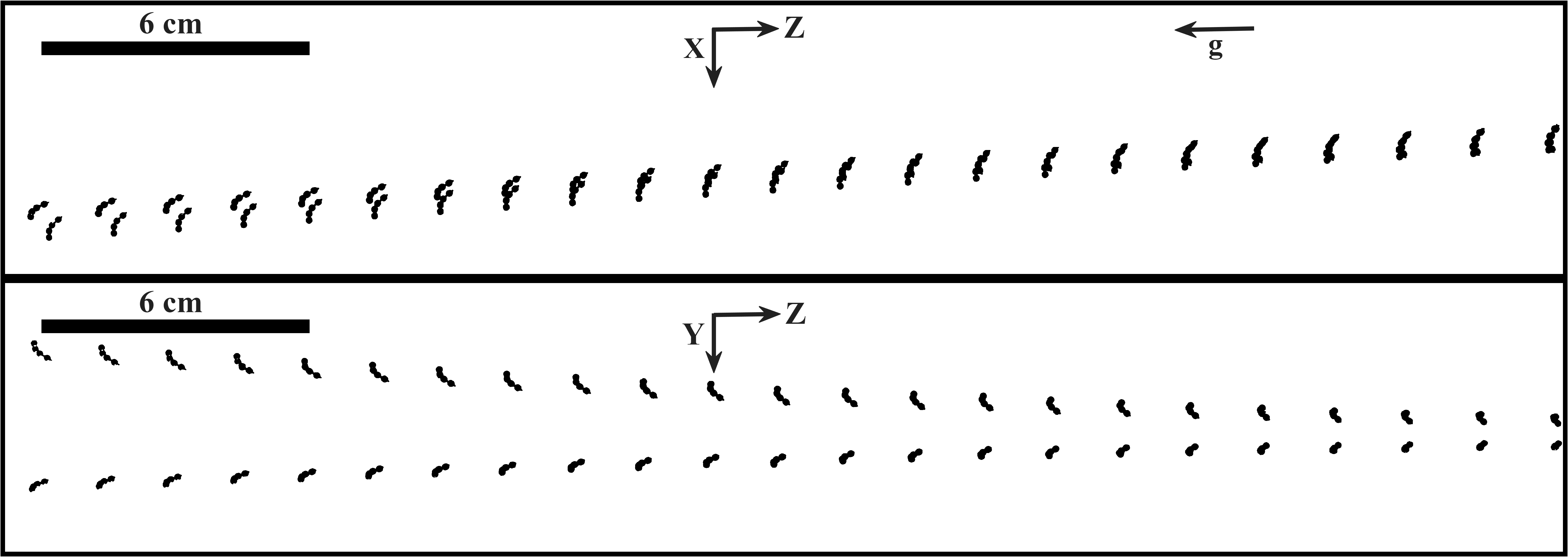}
        \phantomsubcaption
        \label{fig:allmodes:c}
        \vspace{0.25cm} 
    \end{subfigure}
    \begin{subfigure}{0.84\textwidth}
        \centering
    \textbf{(d)} \includegraphics[clip, width=0.99\textwidth]{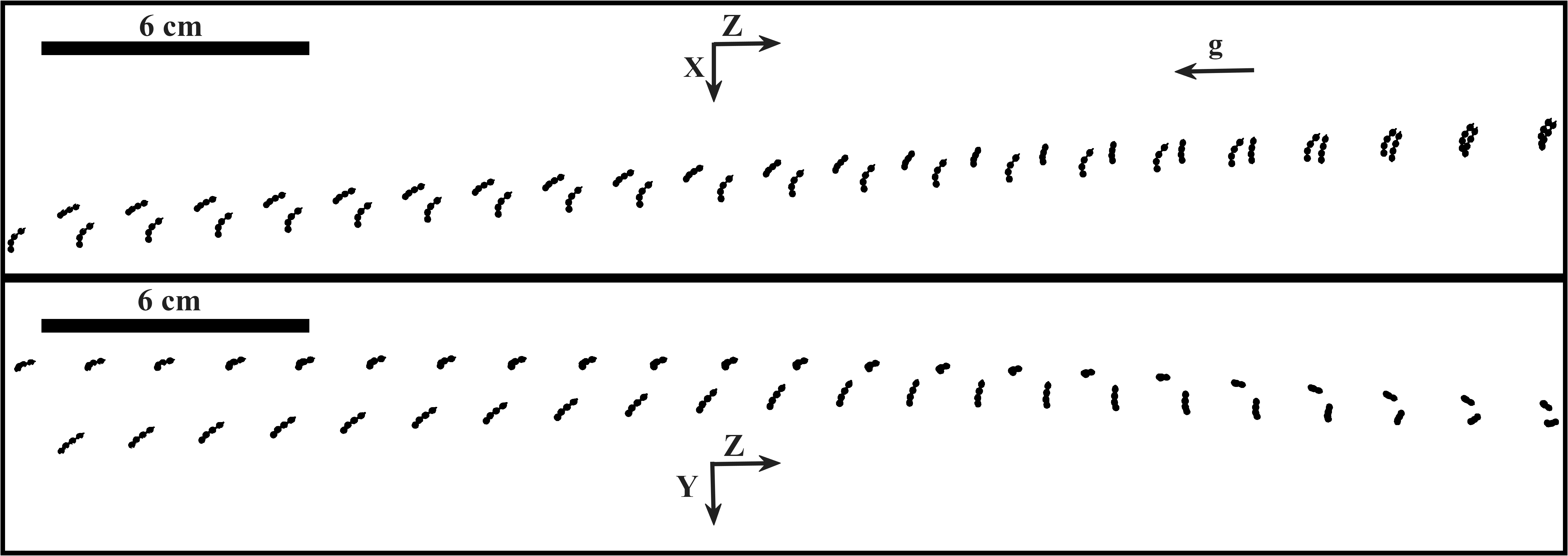}
        \phantomsubcaption
        \label{fig:allmodes:d}
    \end{subfigure}\vspace{-0.2cm}
    \caption{
        Snapshots from 4 experimental trials of two ball chains settling under gravity in a viscous fluid, taken simultaneously by two cameras (top and bottom panels). 
        (a) 3-bead ball chains sedimenting together very close to each other (trial $m_1$), 
        (b) 4-bead ball chains sedimenting together with small oscillations of vertical separation (trial $m_2$), 
        (c) 4-bead ball chains increasing with time their horizontal distance (trial $m_3$), and 
        (d) 4-bead ball chains moving away from each other horizontally and vertically (trial $m_4$). 
        The ball chains move from right to left. 
     }
    \label{fig:allmodes}
\end{figure*}

The main finding is that in most of the experimental trials (in 39 out of 56), the centers of the ball chains remained close to each other, or moved away a little and then came close again, as illustrated in Figs.~\ref{fig:allmodes:a}-\ref{fig:allmodes:b}. 
We referred to this dynamics as `sedimenting together'. 
Within this class, typically, the ball chains stayed relatively close to, or oscillated around, the `aligned' relative configuration \cite{bukowicki_sedimenting_2019}, defined in Eq.~\eqref{aligned}, sometimes with a large oscillation amplitude. 

As described in the previous section, to characterize quantitatively the dynamics of two ball chains moving together, we use the angles $\tilde{\theta}_i$, and the difference $\Delta Z$ of the vertical coordinates $z_{ci}$ of the ball-chain centers, both extracted from the images made by Camera~1 ($xz$ plane). 

\begin{figure*}
    \centering
    \begin{subfigure}{0.32\textwidth}
        \centering
        \includegraphics[width=\textwidth]{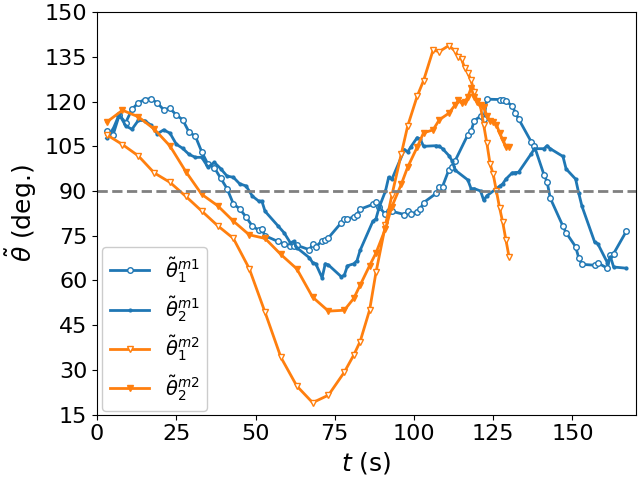}
        \caption{}
        \label{fig:thetavstime:1}
    \end{subfigure}
    \begin{subfigure}{0.32\textwidth}
        \centering
        \includegraphics[width=\textwidth]{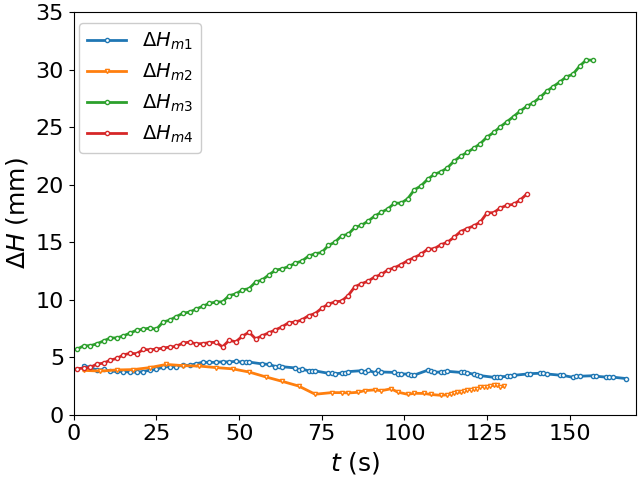}
        \caption{}
        \label{fig:deltaHvstime:1}
    \end{subfigure}
    \begin{subfigure}{0.32\textwidth}
        \centering
        \includegraphics[width=\textwidth]{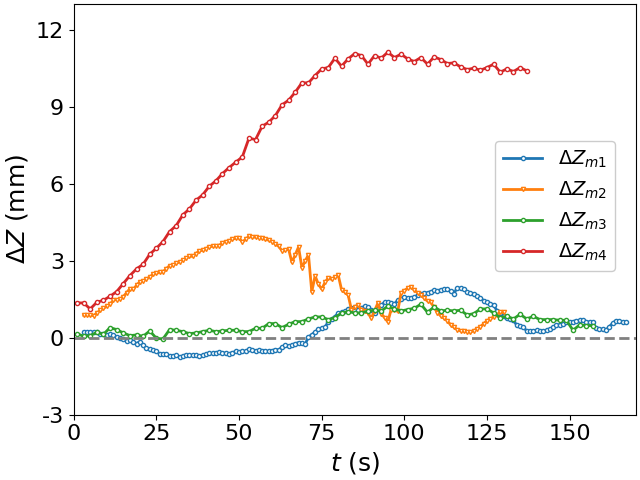}
        \caption{}
        \label{fig:deltaZvstime:1}
    \end{subfigure} 
    \vspace{-0.25cm}
    \caption{
        Time-dependent parameters for the experimental trials, labeled as $m_1$, $m_2$, $m_3$, and $m_4$, the dynamics of which are shown in Figs.~\ref{fig:allmodes:a}-\ref{fig:allmodes:d}, respectively. 
        (a) Inclination angles $\tilde{\theta}_i$ of ball chains $i=1,2$ for trials $m_1$ and $m_2$; 
        (b) horizontal distance $\Delta H$ between the ball-chain centers for all trials $m_1$, $m_2$, $m_3$, and $m_4$; and
        (c) the difference $\Delta Z$ between vertical positions of the ball-chain centers (detected by camera 2) for all trials.
        In trials $m_1$ and $m_2$, the ball chains sediment together, while in trials $m_3$ and $m_4$, they move away from each other.
        }
    \label{fig:exp_param_vs_t}
\end{figure*} 

For 10 trials, while moving together, the two ball chains stay very close to each other. 
They remain approximately parallel to each other, almost within vertical planes parallel to the Camera~1 plane of view. 
Their centers of mass are approximately at the same horizontal level, and at approximately constant horizontal distance $\Delta H$. 
The above properties are summarized as follows: 
  $x_{c1}\!\approx\! x_{c2}$, 
  $y_{c1}\!-\!y_{c2}\!\approx\! \mbox{const}\!\ne\!0$, 
  small $\Delta Z \!=\!z_{c1}\!- \!z_{c2}$, 
  small $\phi_{i}$, 
  $\tilde{\theta}_{i}\!\approx \!\theta_{i}$, and 
  $\tilde{\theta}_{2}$ close to $\tilde{\theta}_{1}$. 
Typically, the angles $\tilde{\theta}_i$ oscillate with time around $90\degree$, almost in phase, with a small amplitude sensitive to the initial orientation and typically slowly decreasing with time. 
The dynamics described above have been observed for the ball chains consisting of 3, 4, 5, and 6 beads. 
A typical example of the snapshots from both cameras is shown in Fig.~\ref{fig:allmodes:a}, and the corresponding time evolution of $\tilde{\theta}_i$, $\Delta H$  and $\Delta Z$ is presented in Fig.~\ref{fig:exp_param_vs_t}.

For 25 trials, the two ball chains move together, but temporarily separate vertically by a small distance, and then come close again. 
For these trials, we observe oscillations with a larger amplitude of both vertical separation between the ball-chain centers, $\Delta Z= z_{c1}-z_{c2}$, and the inclination angles $\tilde{\theta}_i$, as illustrated in Fig.~\ref{fig:allmodes:b} by the snapshots from both cameras for an exemplary trial. 
The time-dependence of $\tilde{\theta}_i$ and $\Delta Z$ is shown in Fig.~\ref{fig:exp_param_vs_t}.
The oscillations of the inclination angles of both ball chains are typically not synchronous; their amplitude decays with time, with the maxima shifted in phase.
The oscillatory behavior is observed across all ball-chain lengths studied, from 3 to 6 beads.

For the ball chains sedimenting together, composed of 5 or 6 beads, in three trials, we observed that they seemed to approach a stationary configuration, with the end-to-end vectors oriented horizontally and parallel to each other, with $\Delta Z \ne 0$ remaining constant over time until the end of the experiment. 
An example is shown in Fig.~\ref{fig:othermodes:b} in Appendix~\ref{appendixA}. 

In some cases, the ball chains rotated around a vertical axis. 
An example is shown in Fig.~\ref{fig:othermodes:a} in Appendix~\ref{appendixA}. 
A small rotation was sometimes observed in addition to the oscillations.

For the other class of the experimental trials (17 out of 56), we observed that the ball chains gradually moved away from each other: 
only horizontally (in 5 trials; an example is shown in  Fig.~\ref{fig:allmodes:c}), 
horizontally and vertically (in 7 trials; an example is shown in  Fig.~\ref{fig:allmodes:d}), or 
only vertically (in 5 trials; and example is shown in Fig.~\ref{fig:othermodes:c} in Appendix~\ref{appendixA}). 
We called such a dynamics `the ball chains at short times moving away from each other'. 
An increase of vertical separation without a horizontal displacement has been observed for more deformed, longer fibers, made of 5 or 6 beads, with the upper fiber less bent than the lower one (and therefore, the lower one sedimenting faster). 

In figure \ref{fig:allmodes:c}, for the largest times, 
a small difference between the vertical coordinates of the ball-chain centers 
is observed by camera 1, while in the image from camera 2, the ball-chain centers have approximately the same vertical coordinates. 
The proper value is given by camera 2, because there is practically no horizontal separation of the ball-chain centers detected by camera 1, and therefore, the calibration factors for the ball-chain positions detected by camera 2 are almost
the same, in agreement with the approximation adopted in this paper.

On the contrary, a precise transformation from pixels to millimeters for the image from camera 1 in Fig.~\ref{fig:allmodes:c} would require a more refined procedure (which would go beyond the scope of the present study). 
The reasoning is as follows. 
Inside the fluid, the height $h'$ of the image at the central vertical line of the container 
is larger than the height $h$ at the front wall, located 100~mm away from the center, owing to a finite camera viewing angle, reduced by light refraction. We measured this difference by placing the ruler inside the tank along its central vertical line, and 
along the central vertical line of the container's inside front wall. 
In this way, we determined the correction factor $h'/h \approx 1.07$. 
A large horizontal separation $\Delta Y \approx 30$ mm between the ball-chain centers, detected by camera~2 at the largest time, results in the difference of the correction factors 
$\approx 0.02$ (scaled linearly as $0.3 \cdot 0.07)$, and therefore, it causes a difference of the calibration factors $w_2-w_1 \approx 0.02 \cdot 0.052$ mm/pixel for the positions of the centers of the ball chains, detected by camera~1 at the largest time. 
The vertical positions of the ball-chain centers in millimeters, measured with respect to the middle horizontal plane, should be the same, $z_{ci}+H/2=w_1(z_{c1,p}+H_p/2)=w_2(z_{c2,p}+H_p/2)$, where $H_p$ = 6720 pixels, and $z_{ci,p}$ is the position of the $i$-th ball-chain center in pixels. 
(Note that $z_{ci}$ and $z_{ci,p}$ are negative and equal to zero at the right edge of the images in Fig.~\ref{fig:allmodes}.)
As seen from Fig.~\ref{fig:allmodes:c}, at the largest time $z_{c1,p} \approx -H_p$, and  $z_{c2,p} \approx -H_p - \Delta Z_p$. Therefore, the value of $-w_2\Delta Z=w_2(z_{c2,p}-z_{c1,p})$ can be estimated as $(w_2-w_1)H_p/2 \approx 3.5$ mm, in agreement with the apparent difference detected by camera 1 at the largest times. 
Analogical reasoning for the experimental trial shown in Fig.~\ref{fig:allmodes:d} supports the conclusion that in this case, $\Delta Z$, given by Eq.~\eqref{definition_of_Deltas}, should also be calculated based on the image from camera~2, as shown in Fig.~\ref{fig:exp_param_vs_t}(c).

Our experiments confirmed that most of the flexible ball chains stayed together, close to an aligned configuration, in agreement with numerical results for moderately stiff elastic fibers, reported in Ref.~\cite{bukowicki_sedimenting_2019} for a different fiber aspect ratio and a different class of initial conditions. 
In the next section, we will use in the numerical simulations a more precise model of hydrodynamic interactions between moderately elastic fibers, and perform a qualitative comparison of the dynamics with the experimental results for flexible ball chains. 
Moreover, in the numerical simulations, we can determine the evolution of a pair of semi-flexible fibers over much longer timescales than in the experiments. 
Therefore, we can investigate if semi-flexible fibers that moved away from each other can again come close, and approach the aligned configuration.

\section{Numerical results}
\label{sec:num}
\vspace{-0.2cm}
\subsection{Parameters and initial conditions}
\label{sec:num:initcond}

To complement the experimental observations, we performed a series of numerical simulations of the elastic filament's dynamics across a range of initial configurations. 
We restrict our study to short fibers composed of $N=4$ very close beads.
To maintain a qualitative consistency with the observed small bending amplitude of the ball chains, we set moderate values of the dimensionless stiffness parameters to $A/\left( d^2Nmg \right) = 0.91809$ and $k/\left( Nmg/d \right) = 14.544$.
These values correspond to the elasto-gravitational number $B \approx 17.7$, which relates gravitational and elastic forces as $B=L_{0}^2Nmg/A$, where $L_0=(N-1)l_0+d$ is the equilibrium length of the fiber. 
To provide context for this choice, highly elastic modes (other than a stable vertical U-shape) have been observed for much larger values, $B\gtrsim 1500-2000$ \cite{saggiorato2015conformations,melikhov2024}.

The multipole expansion applied to evaluate the mobility matrix for elastic filaments was truncated at order $LL=3$. 
This level of approximation has been shown to be well-converged \cite{Cichocki1994,cichocki_lubrication_1999,ekiel-jezewska2009}, and sufficient for capturing the dynamics of similar hydrodynamic systems \citep{shashank2023dynamics}. 
All results are presented in dimensionless units, with lengths scaled by the bead diameter $d$ and time scaled by $\tau_{HM} = \frac{\pi \eta d^2}{Nmg}$.

To qualitatively capture the observed experimental behavior, we initialize the numerical simulations with configurations that reproduce the essential features of the measured dynamics of the ball chains. 
Initially, the elastic filaments are straight and in elastic equilibrium. 
Following the notation established in Section~\ref{sec:theor_num_approach} and illustrated in Fig.~\ref{fig:notation:b}, we define the state of each straight fiber $i\!=\!1,2$ in elastic equilibrium by its center-of-mass position $(x_{ci},y_{ci},z_{ci})$ and its orientation angles $\theta_i$ and $\varphi_i$.

Let us briefly consider the simplest symmetric initial  conditions: two horizontal and parallel straight filaments in elastic equilibrium:
$x_{c2}-x_{c1} = z_{c2}-z_{c1} = 0$, $y_{c2}-y_{c1} \ne 0$, $\theta_{1} = \theta_{2} = 90 \degree$ and $\varphi_{1} = \varphi_{2} = 0$. 
For fibers with a moderate value of the bending stiffness, as in our case, previous studies performed for different symmetric initial conditions \cite{bukowicki_sedimenting_2019, bukowickiPhD} indicate that the fibers relatively fast form slightly tilted wide U-shapes and converge to the aligned configuration, given by Eq.~\eqref{aligned}. 
Later, they attract one another, resulting in a monotonic decrease of their relative horizontal distance $\Delta H$.
We select asymmetric specific starting configurations aiming to qualitatively match the behavior observed in various experimental trials shown in Fig.~\ref{fig:allmodes}.
We consider four distinct initial configurations: 
\vspace{-0.2cm}
\begin{itemize}
    \item
        configuration $S_1$ (Symmetric): 
        the filaments are straight, and slightly inclined with respect to the horizontal in the same direction, and the system maintains reflection symmetry across the mid-plane between the filaments:
        $x_{c2}-x_{c1} =  0$, 
        $y_{c2}-y_{c1} = -4$, 
        $z_{c2}-z_{c1} =  0$, 
        $\theta_{1} = \theta_{2} = 110 \degree$, and 
        $\varphi_{1} = \varphi_{2} = 0$;  \vspace{-0.24cm}
    \item 
        configuration $S_2$ (Weakly asymmetric, perturbed $S_1$): 
        the filaments are straight, and slightly inclined with respect to the horizontal but with a small difference between the inclination angles $\theta_i$, a weak perturbation to their relative positions away from the symmetrical positions and different azimuthal angles:
        $x_{c2}-x_{c1} = 0.6$, 
        $y_{c2}-y_{c1} = 3.3$, 
        $z_{c2}-z_{c1} =  0.05$, 
        $\theta_{1} = 105 \degree$,
        $\theta_{2} = 119 \degree$,  
        $\varphi_{1} = 0$, and
        $\varphi_{2} = -45 \degree$; \vspace{-0.24cm}
    \item 
        configuration $S_3$ (Strongly asymmetric, small vertical separation): 
        the filaments are straight, and severely inclined with respect to the horizontal, with a large difference $\Delta \theta > 90 \degree$ between the inclination angles,
        distinct azimuthal angles $\varphi_{i}$, significant horizontal separation but a small vertical separation:
        $x_{c2}-x_{c1} = -5.5$, 
        $y_{c2}-y_{c1} = -2.6$, 
        $z_{c2}-z_{c1} =  1$, 
        $\theta_{1} = 139 \degree$,
        $\theta_{2} =  41 \degree$,
        $\varphi_{1} = 0$, and
        $\varphi_{2} = 31  \degree$; \vspace{-0.24cm}
    \item 
        configuration $S_4$ (Strongly asymmetric, large vertical separation): 
        the filaments are straight, and severely inclined with respect to the horizontal, with a large difference $\Delta \theta > 90 \degree$ between the inclination angles, 
        distinct azimuthal angles $\varphi_{i}$, significant horizontal separation, and a significant vertical separation:
        $x_{c2}-x_{c1} = -4.7$, 
        $y_{c2}-y_{c1} = -4.8$, 
        $z_{c2}-z_{c1} =  3$, 
        $\theta_{1} = 149 \degree$,
        $\theta_{2} =  51 \degree$,
        $\varphi_{1} = 0$, and
        $\varphi_{2} = 32  \degree$. \vspace{-0.24cm}
\end{itemize}
\begin{figure}[h!]
    \centering
    \begin{subfigure}{0.88\textwidth}
        \centering
    \textbf{(a)} \includegraphics[clip, width=0.95\textwidth]{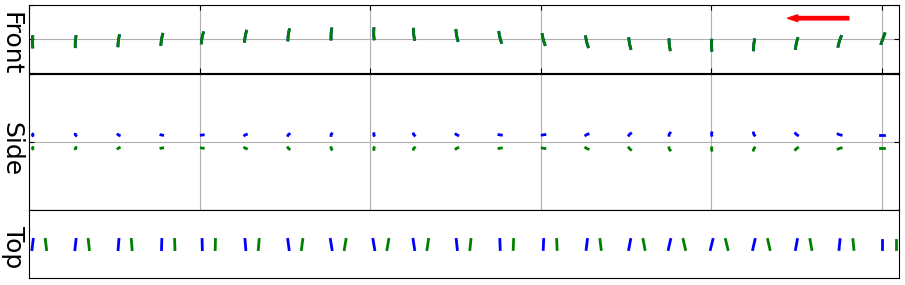}
        \phantomsubcaption
        \label{fig:simul_panels:a}
    \end{subfigure}
    \begin{subfigure}{0.88\textwidth}
        \centering
    \textbf{(b)} \includegraphics[clip, width=0.95\textwidth]{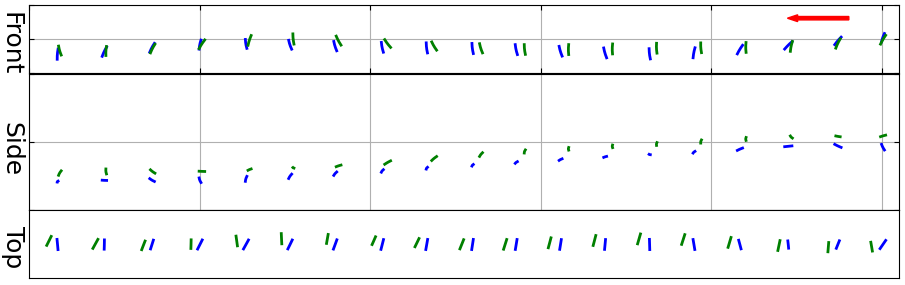}
        \phantomsubcaption
        \label{fig:simul_panels:b}
    \end{subfigure}
    \begin{subfigure}{0.88\textwidth}
        \centering
    \textbf{(c)} \includegraphics[clip, width=0.95\textwidth]{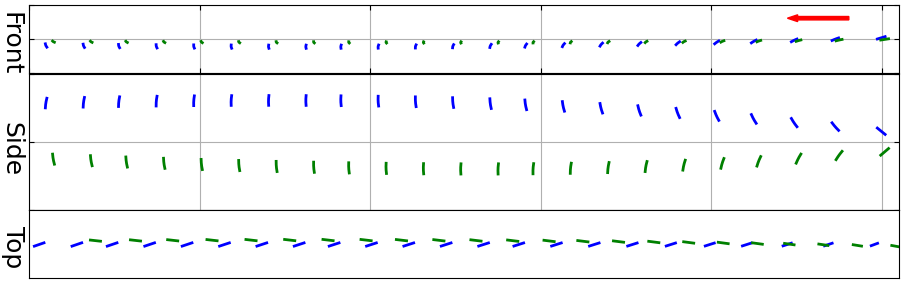}
        \phantomsubcaption
        \label{fig:simul_panels:c}
    \end{subfigure}
    \begin{subfigure}{0.88\textwidth}
        \centering
    \textbf{(d)} \includegraphics[clip, width=0.95\textwidth]{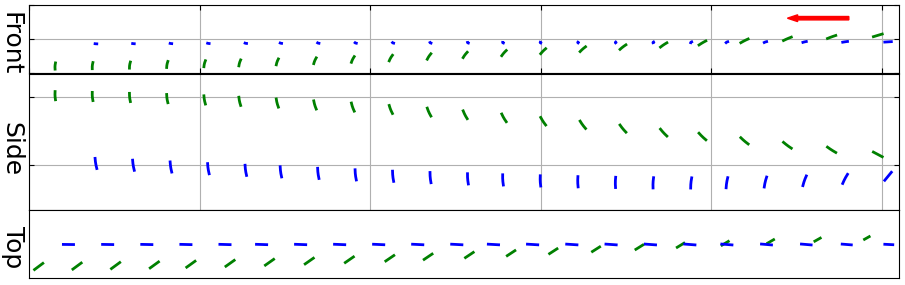}
        \phantomsubcaption
        \label{fig:simul_panels:d}
    \end{subfigure}
    \caption{
        Snapshots from the simulations of two elastic fibers made of 4 beads each, settling under gravity, arranged initially in the configuration: 
        (a) $S_1$, 
        (b) $S_2$, 
        (c) $S_3$, and 
        (d) $S_4$.
        The shown range of vertical center-of-mass positions is equal to $250d$, and it approximately corresponds to the experiments. 
        Snapshots are taken at the time intervals $\Delta t = 68$. 
        Note that in order to reach the level of $z=-250 d$, it takes $\Delta t /\tau_{HM}\approx 1400, 1350, 1550, 1550$ for the fibers in the configurations $S_1$, $S_2$, $S_3$ and $S_4$, respectively.
        Red arrow shows the direction of gravity.
        Videos of the long-time dynamics are provided in Supplemental Material.
    }
    \label{fig:simul_panels}
\end{figure} 

\subsection{Short-time dynamics}
\label{sec:num:short_time} 

To analyze the short-time dynamics, we limit to the range of vertical positions from $z=0$ to $z=-250 d$. Given a bead diameter of $d=1.5$~mm, this distance corresponds approximately to the 350~mm observation window used in our experiments. We observe in the simulations that it takes $t \approx 1350-1550 \tau_{HM}$ to reach the level of $z=-250 d$ by either both or one of the fibers, as illustrated in Fig.~\ref{fig:simul_panels}.

\begin{figure*}
    \centering
    \begin{subfigure}{0.32\textwidth}
        \centering
        \includegraphics[width=\textwidth]{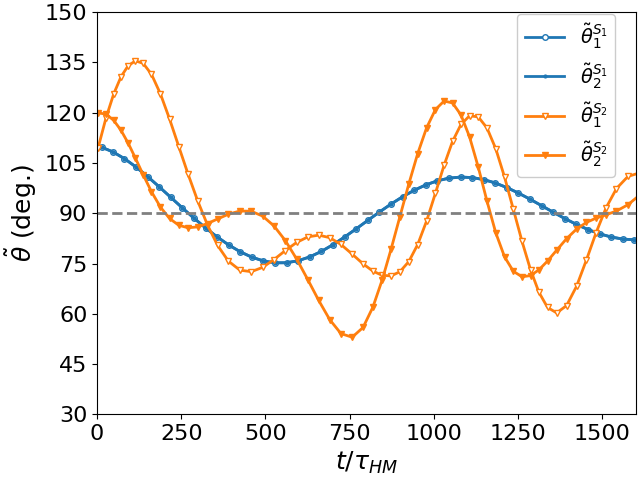}
        \caption{}
    \label{fig:simul:results:a}
    \end{subfigure}
    \begin{subfigure}{0.32\textwidth}
        \centering
        \includegraphics[width=\textwidth]{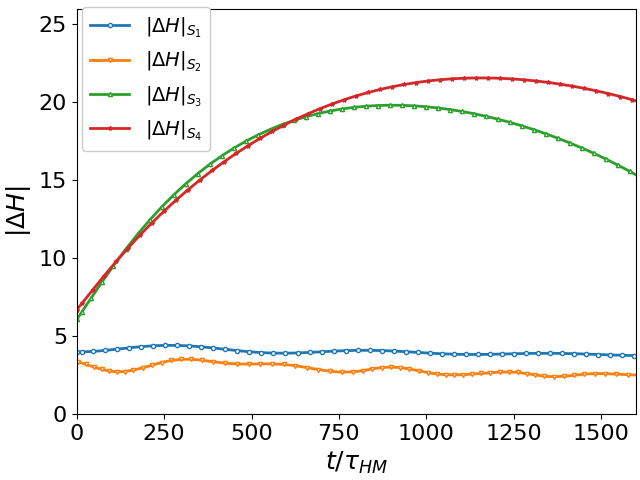}
        \caption{}
    \label{fig:simul:results:b}
    \end{subfigure}
    \begin{subfigure}{0.32\textwidth}
        \centering
        \includegraphics[width=\textwidth]{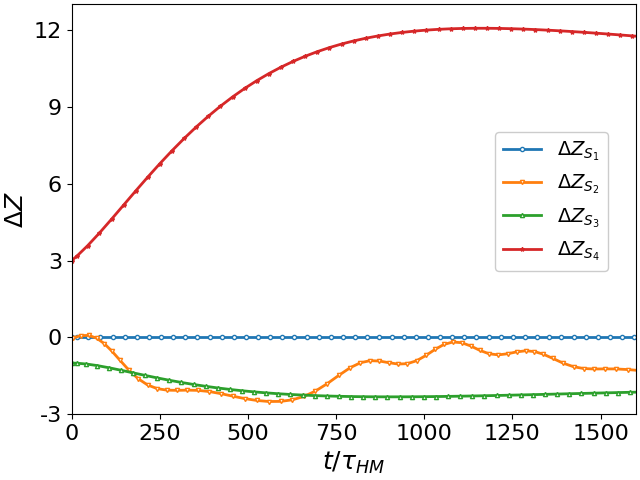}
        \caption{}
    \label{fig:simul:results:c}
    \end{subfigure}
    \vspace{-0.3cm}
    \caption{
        Simulation results for short-time dynamics of elastic fibers. 
        (a) Time-dependence of the orientation angles $\tilde{\theta}_i$ for both fibers $i=1,2$, initially in configurations $S_1$ and $S_2$. 
        (b) Time dependence of the horizontal distance $\Delta H$ and 
        (c) vertical distance $\Delta Z$ between the fiber centers of mass, for the initial configurations $S_1$, $S_2$, $S_3$ and $S_4$. 
    }
    \label{fig:simul:results}
\end{figure*}

In the case of the configuration $S_1$, the symmetric initial conditions lead to the fiber shapes and orientations symmetric with respect to a vertical plane. 
The fibers perform an oscillatory motion of a very small amplitude, belonging to the same class as shown in Ref.~\cite{bukowicki_sedimenting_2019} (see their supplementary video 3) for a different family of initial symmetric configurations. 
In the simulation starting from $S_1$, the orientation angles $\theta_1$ and $\theta_2$ oscillate in phase about the horizontal ($90 \degree$). 
As shown in Fig.~\ref{fig:simul:results:a}, the oscillation amplitude decays over time. 
Throughout this process, as shown in Figs.~\ref{fig:simul:results:b} and~\ref{fig:simul:results:c}, the horizontal separation $\Delta H$ remains nearly constant, and the vertical offset $\Delta Z = 0$. 
The front panel of Fig.~\ref{fig:simul_panels:a} further demonstrates that the centers of mass undergo damped horizontal oscillations about a vertical plane parallel to the side view.

For the configuration $S_2$, where fibers are initially slightly inclined with respect to the horizontal, have different azimuthal angles, and their relative position is slightly different from the symmetric case, later the oscillatory motion is not so regular as for $S_1$. 
The orientation angles $\theta_1$ and $\theta_2$ still oscillate about $90 \degree$ with a small phase shift, and slightly different time-dependent amplitudes (Fig.~\ref{fig:simul:results:a}). 
While the amplitude of oscillation is larger than in the symmetric case, it exhibits a similar temporal decay. 
For this initial configuration, the horizontal distance $\Delta H$ decreases slightly with time (Fig.~\ref{fig:simul:results:b}). 
Notably, the vertical distance $|\Delta Z|$ increases from zero to a maximum of approximately $3d$ at $t \approx 600 \tau_{HM}$, before returning to $\Delta Z \approx0$ at $t \approx 1100 \tau_{HM}$ (see Fig.~\ref{fig:simul:results:c}). 
Similar to the symmetric case, the centers of mass exhibit oscillatory motion about the vertical plane parallel to the camera side view, observed in the front panel of Fig.~\ref{fig:simul_panels:b}.
In addition, there is also a slight side movement of both fibers relative to a plane parallel to the front view, observed in the side panel of Fig.~\ref{fig:simul_panels:b}.

Properties of the simulation dynamics, captured in Fig.~\ref{fig:simul:results} for the initial conditions $S_1$ and $S_2$, qualitatively mirror the experimental behavior of the ball chains that sediment close to each other, i.e., trials $m_1$ and $m_2$ in Fig.~\ref{fig:exp_param_vs_t}: 
the inclination angles $\theta_{i}$ oscillate nearly in phase with decaying amplitudes, 
the horizontal separation $\Delta H$ remains largely stable, and,
for trial $m_2$, the vertical separation $|\Delta Z|$ reaches a local tiny maximum during the first third of the observed trajectory.

To induce substantial horizontal or vertical separations of the elastic filaments in the simulations, the initial configurations $S_3$ and $S_4$ were modified significantly from the symmetric case.
First, the initial horizontal separations between the fibers were increased to $6.1d$ and $6.7d$ in the configurations $S_3$ and $S_4$, compared to $4.0d$ and $3.35d$ for $S_1$ and $S_2$. 
The initial vertical separations were correspondingly larger ($1d$ and $3d$, versus $0$ and $0.05d$). 
In $S_3$ and $S_4$, the fibers were given much steeper initial inclinations, with $|\theta_i-90\degree|$ ranging from $40\degree - 60 \degree$ compared to $15\degree - 30 \degree$ for $S_1$ and $S_2$, and $\Delta \theta$ around $100\degree$, compared to $0\degree-15\degree$.

As established in previous studies \cite{bukowicki_sedimenting_2019, bukowickiPhD, melikhov2024}, a single, relatively short, moderately elastic fiber sedimenting at an initial inclination will translate sideways while settling down before eventually adopting a bent shape and a stable very wide U-shaped orientation. 
For larger distances between fibers, this effect is larger than the mutual hydrodynamic interactions of both fibers.
Therefore, initiating the fibers at larger inclination angles prolongs this horizontal translation during the early stages of sedimentation, enhancing the resulting horizontal  separation. 
Moreover, the difference between $|\theta_1-90\degree|$ and $|\theta_2-90\degree|$ causes a difference between the vertical velocity component of both filaments, resulting in their vertical separation, as illustrated in Fig.~\ref{fig:simul_panels:d}. 
This kinematic behavior is clearly seen in the side panels of Figs.~\ref{fig:simul_panels:c} and \ref{fig:simul_panels:d}, and is quantitatively confirmed in Figs.~\ref{fig:simul:results:b} and \ref{fig:simul:results:c}.

For both of the strongly asymmetric configurations $S_3$ and $S_4$, the horizontal separation $\Delta H$ reaches a maximum of $\approx 20d$ around $t=1000 \tau_{HM}$. 
In contrast, the vertical separation $|\Delta Z|$ increases significantly only for the configuration $S_4$ that has a larger initial vertical offset, and a large difference between $|\theta_1-90\degree|$ and $|\theta_2-90\degree|$, reaching a maximum of $\approx 12d$ around $t=1000 \tau_{HM}$. 
In the short-time range, these numerical dynamics are qualitatively similar to the experimental observations of the ball chains that move away from each other at short times, i.e., specifically trials $m_3$ and $m_4$ in Fig.~\ref{fig:allmodes}, with parameters reported in Fig.~\ref{fig:exp_param_vs_t}.

The numerical simulations starting from $S_3$ and $S_4$ show the existence of maxima of horizontal or horizontal-and-vertical time-dependent separations between the elastic fibers. Therefore, in the next section, we will explore what happens later.

\subsection{Long-time dynamics}
\label{sec:num:long_time}

The long-time dynamics of elastic fibers initially in a symmetric configuration of straight horizontal fibers with $\phi_1=45\degree,\;\phi_2=135\degree$ and $y_{c1}=y_{c2}$ have been reported in Ref.~\cite{bukowicki_sedimenting_2019}. It was shown there numerically using the Rotne-Prager approximation that stiffer fibers approach the terminal configuration in two stages. First, they perform oscillations of their orientation angles $\theta_i$ and $\phi_i$, with $i=1,2$, and relative position. The amplitude $A(n)$ of the oscillations of $\theta_i$ was shown to decay exponentially with the number of the oscillation $n$. At the end of the first stage, oscillations are practically damped, and the fibers are very close to the aligned configuration.
In the second stage, the fibers remain very close to the aligned configuration, but translate horizontally towards touching their end beads \cite{bukowicki_dynamics_2015}. 

In this section, we will present the long-time dynamics for different initial conditions.
From now on, we will use the orientation angles $\theta_i$ (rather than $\tilde{\theta}_i$ determined in the previous subsection).
\begin{figure*}
    \centering
    \begin{subfigure}{0.32\textwidth}
        \centering
        \includegraphics[width=\textwidth]{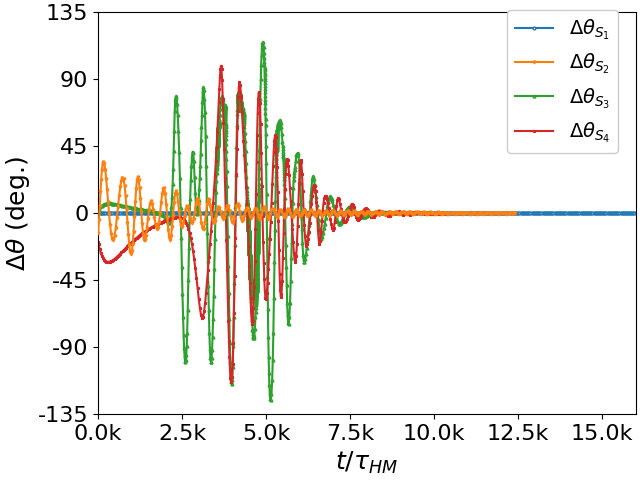}
        \caption{}
        \label{fig:simul:long_time_results:a}
    \end{subfigure}
    \begin{subfigure}{0.32\textwidth}
        \centering
        \includegraphics[width=\textwidth]{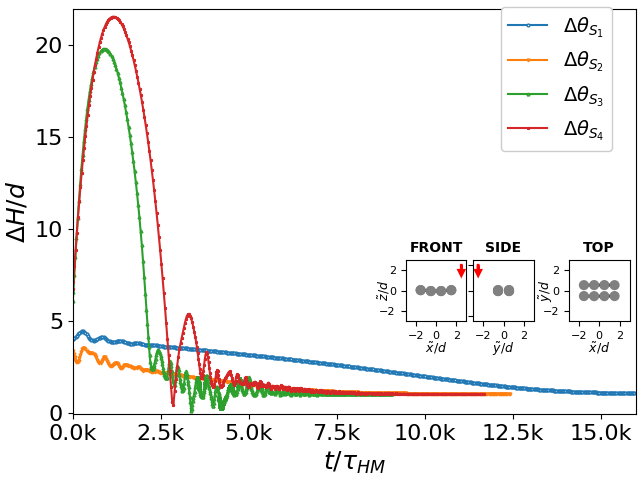}
        \caption{}
        \label{fig:simul:long_time_results:b}
    \end{subfigure}
    \begin{subfigure}{0.32\textwidth}
        \centering
        \includegraphics[width=\textwidth]{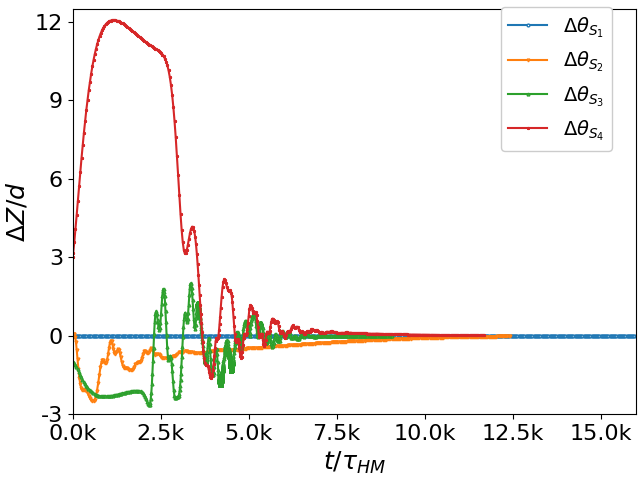}
        \caption{}
        \label{fig:simul:long_time_results:c}
    \end{subfigure}\vspace{-0.2cm}
    \caption{
        Simulation results for long-time dynamics of elastic fibers with the initial configurations $S_1$, $S_2$, $S_3$ and $S_4$ (videos are provided in Supplemental Material).
        (a) Difference $\Delta \theta(t)$ between the orientation angles of the fibers; 
        (b) horizontal distance $\Delta H(t)$ and 
        (c) vertical distance $\Delta Z(t)$ between the fiber centers of mass.
        Inset in (b) illustrates that at the end of each simulation, the elastic filaments approached approximately the same aligned relative configuration, defined in Eq.~\eqref{aligned}, with $|\bm{r}_{c1}\!-\!\bm{r}_{c2}|/d$ very close to unity (the reference frame was rotated, and its center was shifted).
        Red arrow shows the direction of gravity.
    }
    \label{fig:simul:long_time_results}
\end{figure*} 

To investigate the long-term sedimentation dynamics, we extended the simulations illustrated in Fig.~\ref{fig:simul:results} until the filaments reached a steady relative configuration. 
For the initial conditions $S_1-S_4$, even though the short-time dynamics of elastic fibers looked different, the fibers eventually approached approximately the same aligned configuration, with the end beads almost touching each other. 
The simulations were terminated when the fibers were nearly in contact, i.e., when the smallest distance between the centers of the end beads from different chains became as small as $1.005d$.
This proximity was achieved at different times: $t/\tau_{HM}=20000$, $13000$, $9100$, and $11700 $ for the initial configurations $S_1$, $S_2$, $S_3$, and $S_4$, respectively. 
The terminal relative configuration adopted by the fibers for the initial conditions $S_1-S_4$ is shown in the inset of Fig.~\ref{fig:simul:long_time_results:b}. 
Minor deviations from the symmetry are displayed in Fig.~SM2 in Supplemental Material.

The basic features of the system evolution for the initial conditions $S_1-S_4$ are shown in Fig.~\ref{fig:simul:long_time_results}. 
Movies and additional results are provided in Supplemental Material.

Starting from a symmetric initial configuration $S_1$, we observe a similar behavior as in Ref.~\cite{bukowicki_sedimenting_2019} for a different symmetric initial arrangement. Each fiber $i$ undergoes damped oscillations of the orientation angles $\theta_i$ \& $\phi_i$ (see Movie S1 and Fig.~SM1 in Supplemental Material). In addition, the fibers slowly approach each other horizontally, as visible in Fig.~\ref{fig:simul:long_time_results:b}. The decay time is much shorter than the approach time. The fibers' relative positions and orientations are very well approximated by the aligned configuration with negligible oscillations even when they are still separated by a quite large horizontal distance, which slowly decreases until $t \approx 10000 \,\tau_{HM}$ (see Movie~S1 in Supplemental Material).

Similarly, the fibers initially at $S_2$ configuration from the beginning exhibit decaying orientational oscillations (see Movie~S2 and Fig.~SM1 in Supplemental Material).
However, for $S_2$, the decay time is comparable to the time of the horizontal approach, but shorter than the time of the vertical approach, as shown in Figs.~\ref{fig:simul:long_time_results} and SM1. As a result, in the terminal configuration, the dot products in Eq.~\eqref{aligned} are non-zero but negligible.

The initial conditions $S_3$ and $S_4$ differ significantly from the aligned configuration. In their early evolution, the fibers move away from each other vertically and horizontally, without any oscillations, as visible in Figs.~\ref{fig:simul:long_time_results:b}-\ref{fig:simul:long_time_results:c} and supplementary movies S3 and S4.

For $S_4$, the horizontal and vertical distances between the fibers,  $\Delta H$ and $|\Delta Z|$, significantly increase with time. The fibers seem to separate permanently. However, for 
$\Delta H \approx 22d$ and $|\Delta Z| \approx 12d$, they start to approach each other. When they are sufficiently close, their orientation angles and the relative position start to oscillate with a rapidly increased amplitude.

A similar behavior is observed for $S_3$, but with a much smaller maximum value of $|\Delta Z|$. 
For both $S_3$ and $S_4$, the oscillations start when the fibers are relatively close to each other, as visible in Figs.~\ref{fig:simul:long_time_results:b}-\ref{fig:simul:long_time_results:c} and supplementary movies S3 \& S4. 
The fibers slowly approach the aligned configuration, with damped oscillations of the relative position and of the orientation angles $\theta_i$. 
At the terminal configuration, the fibers are almost aligned, with the amplitude of the last oscillation of $\theta_1$ around $5\degree$ for $S_3$ and $2\degree$ for $S_4$.
While the deviations from perfect alignment in the terminal configurations for $S_3$ and $S_4$ are slightly larger than for $S_1$ and $S_2$, they remain within a small neighborhood of the aligned state, as can be seen in Fig.~SM2 in Supplemental Material. 
The terminal states for $S_3$ and $S_4$ are well-approximated by the same inset in Fig.~\ref{fig:simul:long_time_results:b}.

Figures~\ref{fig:simul:long_time_results:b} and ~\ref{fig:simul:long_time_results:c} illustrate how differently fibers approach the aligned configuration, depending on the initial conditions.
Notably, for the initial mirror symmetry (including $\Delta \theta=0$ and $\Delta Z=0$) in $S_1$, the horizontal separation $\Delta H$ relaxes toward $1d$ more slowly than in the asymmetric cases. 
Consequently, the center-of-mass distance remains significant ($\sim 2d$) even for $t \approx 7500\tau_{HM}$, when the oscillations are practically damped, for $S_1$ as well as for $S_2-S_4$. 
Also, we observe that the decrease of $|\Delta Z|$ for the slightly asymmetric initial configuration $S_2$ is significantly slower than for the other initial conditions.

Fig.~\ref{fig:simul:long_time_results} also illustrates that for $S_3$ and $S_4$, the oscillations of $\theta_i$ couple with each other and with the oscillations of $\Delta Z$ and $\Delta H$ in a different way than for $S_1$ and $S_2$. 
In comparison with $S_1$ and $S_2$, for $S_3$ and $S_4$ the amplitude $A_{\theta_i}$ of oscillations of $\theta_i$ is larger, and the amplitude of the oscillations of $\Delta \theta$ is almost twice larger than $A_{\theta_i}$ (compare with Fig.~SM1). 
The oscillations of the fiber orientation are out of phase, which is clearly visible in supplementary movies S3 and S4.
Furthermore, for $S_3$ and $S_4$, $\Delta H$ reaches near zero occasionally, indicating intervals where one fiber passes directly above the other. 
Moreover, there appear oscillations of the fibers' relative vertical position $\Delta Z$ around zero: a fiber goes above and below the other one in a repeatable pattern, and decaying amplitude of the oscillations.

Despite the system exploring large separations, i.e., reaching horizontal distances of
$\Delta H \gtrsim 20d$ for $S_3$ and $S_4$ and vertical offsets as large as $\Delta Z \approx 12d$ for $S_4$, these values eventually decay to $\approx 1d$ and $0$, respectively. 
Our results demonstrate that the hydrodynamic interactions between the filaments are inherently attractive, leading all investigated initial configurations to converge to a closely spaced, aligned terminal state.

\section{Conclusions}
\label{sec:concl}

This paper is focused on the dynamics of two short, identical, moderately flexible fibers settling under gravity close to each other in a viscous fluid at the Reynolds number much smaller than unity. The basic question is how the attractive aligned configuration of close fibers (described by Eq.~\eqref{aligned} and the requirement of very close proximity of both fibers) is reached from different initial conditions, and whether it is possible that fibers come to this configuration if separated by a large horizontal or vertical distance. 

To investigate this problem, we started from in general non-symmetric configurations, to supplement the results from Ref.~\cite{bukowicki_sedimenting_2019} where a class of symmetric initial positions of elastic fibers with rather large bending stiffness was shown to converge to the aligned orientations and close positions. 
In our experiments, flexible fibers were modeled as chains of beads with restricted bending angles, and their evolution was recorded over a short time.
We also performed long-time numerical simulations of the dynamics of elastic filaments with a relatively large bending stiffness. 

In the experiments, the ball chains were initially close to the aligned configuration with a moderate distance between straight, horizontal and parallel ball chains, with approximately random deviations. Most of the ball chains exhibit relatively small oscillations of their orientations and positions. But some of them moved away from each other horizontally or vertically. To trace the dynamics for longer times, we performed simulations of moderately elastic filaments. For shorter times, we observed similar behavior as in the experiments. We provided examples illustrating that elastic filaments can quite fast separate from each other significantly, but after a long time, they come back to close proximity, with similarly damped oscillations of their orientations and relative positions around the aligned state. 

The existence of the attractive relative configuration of two very close elastic fibers, with a large basin of attraction, can have many useful biological, medical, and industrial applications, including clustering of micro-organisms or other micro-objects, creating ordered suspensions, transferring information between cells, and triggering chemical reactions. 

Our findings indicate the complexity of the dynamics of two flexible fibers settling under gravity in a viscous fluid. Such a system has a large number of degrees of freedom and has not been fully analyzed yet. It is known that there also exist other attracting states or orbits \cite{saggiorato2015conformations, bukowicki_sedimenting_2019,bukowickiPhD}, and the dynamics still need to be studied. The applied methods and obtained results for single fibers might be useful \cite{melikhov2024,fox2024}. Moreover, it would be worthwhile to investigate attracting configurations or orbits for the dynamics of sedimenting elastic sheets \cite{yu2023dynamics,yu2024free}, e.g., in connection with possible applications for graphene flakes~\cite{salussolia2022simulation}.

\section*{Acknowledgments}
    The authors are grateful to Dr.~H.~J.~Shashank for sharing the MATLAB code and for his insightful discussions of the experiments.
    This work was supported in part by the National Science Center, Poland, under grant UMO-2021/41/B/ST8/04474. 
    H.~N.~M. would like to express his gratitude to his late father, K.~V.~Nagaraj Mirajkar, for instilling in him the values of education and perseverance; his memory remains a constant source of strength.

\section*{Data Availability Statement}
    The data that support the findings of this study are openly available in RepOD -- Repository for Open Data at \url{https://doi.org/10.18150/HEKYSR}.

\appendix

\section{Method used to extract the ball-chain geometric parameters from 2D images}
\label{appendix_bead_detection}

\subsection{Bead detection}
\label{sec:exp_meth:pa:bead_detect}

For each processed frame $f$ from Camera~1, bead images are modeled as approximately circular objects of nearly uniform size in the image plane
The bead-detection procedure is applied to the corresponding binary image and returns a set of candidate bead centers,
\begin{equation*}
    \mathcal{C}^{(f)}=
    \left\{
    \left(x_n^{(f)},z_n^{(f)}\right)
    \;\big|\;
    n=1,2,\dots,N_f
    \right\},
\end{equation*}
where $N_f$ is the number of detected beads in frame $f$. In this convention, $z$ increases to the right in the image plane and $x$ increases downward (see Fig.~\ref{fig:allmodes}).

The primary detection step uses the gradient-based circle detector implemented in OpenCV as \texttt{HoughCircles}. In the present implementation, the search is restricted to a narrow interval of radii centered around a common expected bead radius $R=13$ pixels, so that all detected beads are effectively treated as having the same nominal diameter up to a small tolerance. In the primary detection step, the allowed radius range is $R\pm 2$ pixels. The output of this step is therefore a set of detected circles with centers $\left(x_n^{(f)},z_n^{(f)}\right)$ and radii close to $R$. Only the center coordinates are retained for further analysis.

If this first Hough-transform pass yields only a very small number of detections, a second Hough-transform pass with modified detection parameters is attempted. In this second pass, the allowed radius range is widened to $R\pm 3$ pixels, and the thresholds controlling the gradient-based edge selection and circle acceptance are also adjusted. Therefore, the second pass does not amount merely to enlarging the admissible radius interval, but to a more general modification of the Hough-based detection criteria. This is intended to improve robustness in difficult frames while still avoiding overly permissive detections.

If the result is still insufficient after these two Hough-based attempts, a conservative contour-based fallback procedure is used. In that case, connected contours in the binary image are filtered using geometric criteria based on their area and circularity, consistent with the expected bead size. More precisely, if $A$ denotes the contour area and $P$ its perimeter, then only contours with area lying in a prescribed interval around the reference disk area $\pi R^2$ are accepted; in the implementation used here, this interval is $0.70\,\pi R^2 \le A \le 1.30\,\pi R^2$. In addition, the contour circularity is quantified by the standard shape factor
\begin{equation*}
    \chi=\frac{4\pi A}{P^2},
\end{equation*}
which is equal to $1$ for a perfect circle. Only contours with sufficiently large circularity are retained; in the present implementation, the condition $\chi \ge 0.75$ is imposed. For each accepted contour, the bead center is taken to be the centroid of the contour. Thus, depending on the detection branch, the bead center is defined either as the center of the detected circle or as the centroid of the accepted contour. In practice, this contour-based fallback branch was used only rarely.

To suppress duplicate detections, candidate centers located within a prescribed small distance of one another are merged into a single detection. In the implementation used here, this merging is performed whenever the distance between two detected centers does not exceed $0.5R$, which for $R=13$ pixels corresponds to about $6$ pixels. The resulting set of bead centers is then used in the subsequent chain-identification step.

\subsection{User-guided chain identification}
\label{sec:exp_meth:pa:user_guide}

After automatic bead detection, an interactive zoomed preview of the frame is presented to the user. On this basis, the user manually specifies two ordered lists of detected beads corresponding to Chain~1 and Chain~2. The ordering is imposed from the designated start of a chain to its designated end. If some interior beads are not reliably detected, the user may specify only the first and last visible beads, so that the chain still remains identifiable through its endpoints.

As a result, for each frame $f$ and each chain $i=1,2$, one obtains an ordered list of bead centers,
\begin{equation*}
    \mathcal{C}^{(f)}_{i}
    =
    \left\{
    \left(x_{i, \,n}^{(f)},z_{i,\, n}^{(f)}\right)
    \;\big|\;
    n=1,\dots,M_i^{(f)}
    \right\},
\end{equation*}
where $2 \le M_i^{(f)}\!\le N$ is the number of beads assigned to chain $i$ in frame $f$.

Frames can also be skipped at this stage if the automatic detection is insufficient for a reliable manual assignment of the chains. Therefore, only frames for which the user accepted a consistent bead assignment are included in the subsequent geometric analysis.

\subsection{Geometric descriptors of the chains}
\label{sec:exp_meth:pa:geom_descr}

For each chain $i$ in frame $f$, the center-of-mass position is defined as the arithmetic mean of the selected bead-center coordinates, \vspace{-0.2cm}
\begin{equation*}
    \left(x_{ci}^{(f)},z_{ci}^{(f)}\right)
    =
    \frac{1}{M_i^{(f)}}
    \sum_{n=1}^{M_i^{(f)}}
    \left(x_{i,\,n}^{(f)},z_{i,\,n}^{(f)}\right).
\end{equation*}
Since all beads are treated as identical, no additional weighting is introduced.

The end-to-end vector of chain $i$ in frame $f$ is defined using the first and last beads in the user-provided ordered list,
\begin{equation*}
    \tilde{\mathbf v}_i^{(f)}
    =
    \left(
    x_{i,\,M_i^{(f)}}^{(f)}-x_{i,\,1}^{(f)},
    \;
    z_{i,\,M_i^{(f)}}^{(f)}-z_{i,\,1}^{(f)}
    \right).
\end{equation*}
Its magnitude,
\begin{equation*}
    \left\|
    \tilde{\mathbf v}_i^{(f)}
    \right\|
    =
    \sqrt{
    \left(x_{i,\,M_i^{(f)}}^{(f)}-x_{i,\,1}^{(f)}\right)^2
    +
    \left(z_{i,\,M_i^{(f)}}^{(f)}-z_{i,\,1}^{(f)}\right)^2
    },
\end{equation*}
is the end-to-end distance of the chain in the image plane. This definition remains well posed even when only the two end beads are specified.

As in the main text, the angle $\tilde{\theta}_i^{(f)}$ is measured from the positive $z$-axis to the projected end-to-end vector $\tilde{\mathbf v}_i^{(f)}$. Denoting
\begin{equation*}
    \Delta x_i^{(f)} = x_{i,\,M_i^{(f)}}^{(f)}-x_{i,\,1}^{(f)},
    \qquad
    \Delta z_i^{(f)} = z_{i,\,M_i^{(f)}}^{(f)}-z_{i,\,1}^{(f)},
\end{equation*}
the corresponding angle is determined from the endpoint vector. Its numerical value therefore depends on the ordering of the endpoints selected by the user. If the chain orientation is treated as undirected, the angle is understood modulo $\pi$; otherwise it is treated as a directed angle.

Finally, from the chain centers of mass, one computes the relative displacement components in the Camera~1 plane,
\begin{equation*}
    \Delta X^{(f)} = x_{c1}^{(f)}-x_{c2}^{(f)},
    \qquad
    \Delta Z^{(f)} = z_{c1}^{(f)}-z_{c2}^{(f)},
\end{equation*}
as well as the planar distance between the two centers of mass,
\begin{equation*}
    D_{xz}^{(f)}
    =
    \sqrt{
    \left(\Delta X^{(f)}\right)^2
    +
    \left(\Delta Z^{(f)}\right)^2
    }.
\end{equation*}
These quantities characterize the relative position of the two chains in each processed frame. 
In the main text, the full horizontal separation $\Delta H$ is obtained only after combining the information from Camera~1 and Camera~2, according to Eq.~\eqref{definition_of_Deltas}.

\section{Additional experimental trials}
\label{appendixA}
\begin{figure}[h!]
   \centering
   \begin{subfigure}{0.82\textwidth}
        \centering
    (a) \includegraphics[clip, width=0.99\textwidth]{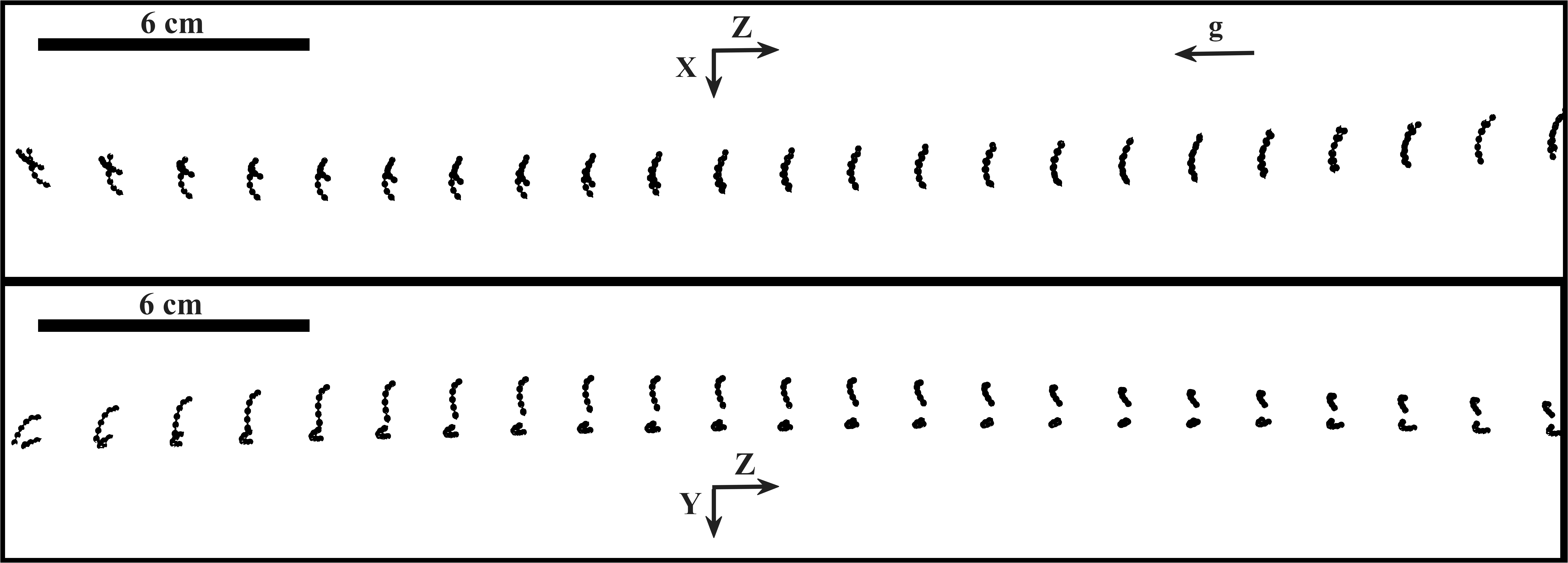}
        \phantomsubcaption
        \label{fig:othermodes:a}
        \vspace{0.2cm}
    \end{subfigure}
    \begin{subfigure}{0.82\textwidth}
        \centering
    (b) \includegraphics[clip, width=0.99\textwidth]{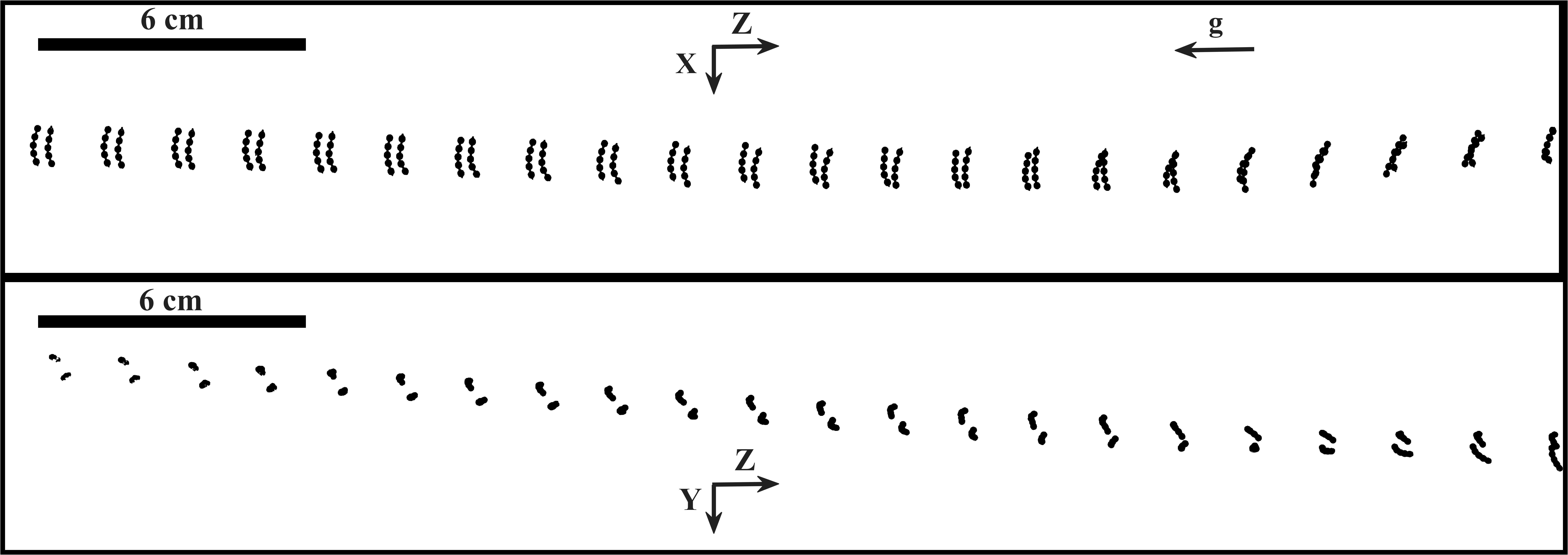}
        \phantomsubcaption
        \label{fig:othermodes:b}
        \vspace{0.2cm}
    \end{subfigure}
    \begin{subfigure}{0.82\textwidth}
        \centering
    (c) \includegraphics[clip, width=0.99\textwidth]{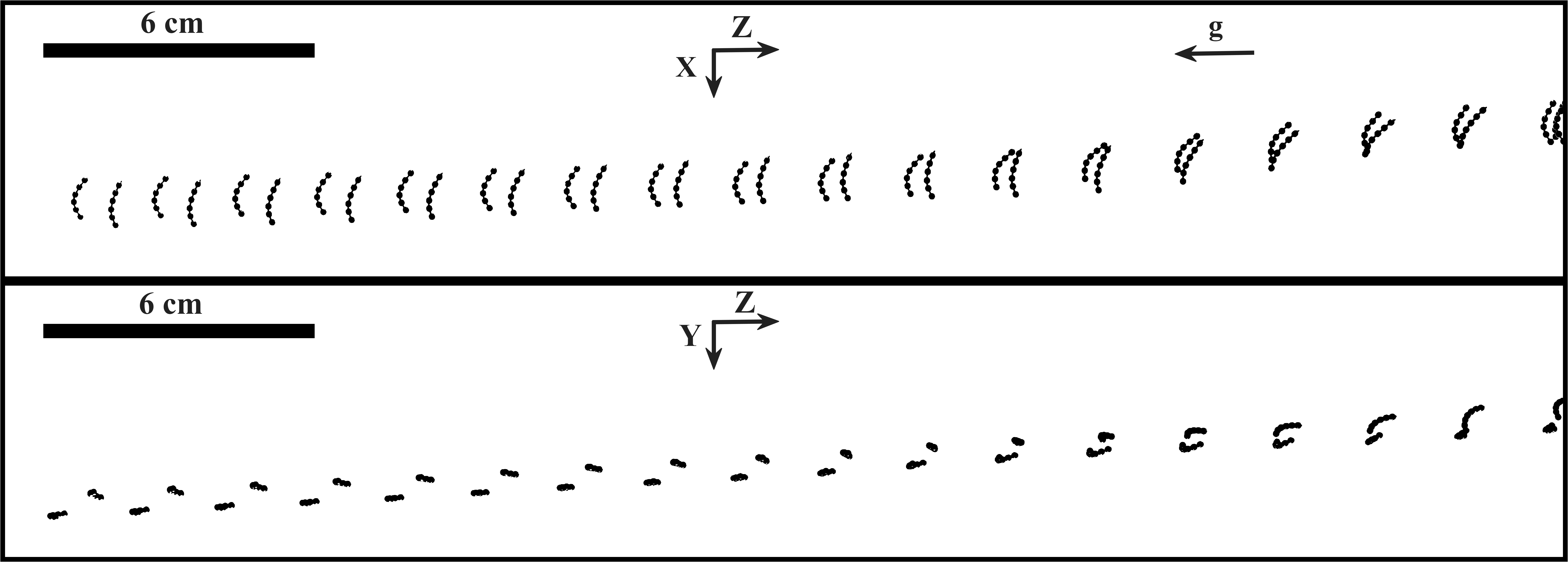}
        \phantomsubcaption
        \label{fig:othermodes:c}
    \end{subfigure}
    \caption{
        Snapshots from 3 experimental trials of two ball chains settling under gravity in a viscous fluid: 
        (a) One of the 6-bead ball chains rotates along $z$ (trial $m_5$);
        (b) 5-bead ball chains experience a prolonged constant vertical separation (trial $m_6$); and
        (c) 6-bead ball chains increase with time their vertical separation, and the bottom ball chain is more curved than the top one (trial $m_7$). 
        Snapshots taken simultaneously by two cameras (top and bottom panels) located at the same level, and with perpendicular horizontal lines of sight. 
        Gravity points left, and the particles move from right to left.
    }
    \label{fig:othermodes}
\end{figure}
As described in Sec.~\ref{sec:res_discuss}, the main modes of the short-time dynamics of two ball chains, observed in our experiments correspond to moving together, with smaller or larger damped oscillations of $\tilde{\theta}_i$, $\phi_i$, $\Delta X$, and $\Delta Z$ around zero, and almost constant $\Delta Y$ (as illustrated in Figs~\ref{fig:allmodes:a}-\ref{fig:allmodes:b}). 
We detected also the ball chains moving away from each other horizontally or horizontally and vertically (as shown in Figs.~\ref{fig:allmodes:c}-\ref{fig:allmodes:d}). 

In this Appendix, we present examples of a different behavior seen in a small number of the experimental trials. 
In Fig.~\ref{fig:othermodes:a}, an example of rotation around the gravitational axis $z$ is shown. 
Fig.~\ref{fig:othermodes:b} illustrates that in some cases, the ball-chain end-to-end vectors ${\bf v}_i$ tend to be parallel to each other and horizontal, but not in the same horizontal plane (with a constant in time $\Delta Z \ne 0$ and $\Delta X$ close to zero). 
In Fig.~\ref{fig:othermodes:c}, we present the ball chains with $\Delta Z$  increasing with time; the lower ball chain is more curved than the upper one and therefore, it sediments faster than the upper one. Within the last stage of this experimental trial, the end-to-end vectors are approximately parallel to each other and horizontal.

\bibliographystyle{unsrtnat}

\bibliography{references.bib}

\end{document}